\documentclass[aps,pre,superscriptaddress,showpacs,amsmath,amssymb,reprint,nofootinbib]{revtex4-1}
\usepackage{amsmath}
\usepackage{amsfonts}
\usepackage{amsthm}
\usepackage{graphics}
\usepackage[draft]{graphicx}
\usepackage{amssymb}
\usepackage{mathtools}
\usepackage{caption,subcaption}
\usepackage{color}
\usepackage{url}
\usepackage[abs]{overpic}
\usepackage[usenames,dvipsnames]{xcolor}
\usepackage[normalem]{ulem}

\def\kmax{k_{\rm max}}

\DeclareMathOperator{\sgn}{sgn}
\DeclareMathOperator{\Tr}{Tr}
\DeclareMathOperator{\Var}{Var}
\DeclareMathOperator{\Exp}{Exp}

\graphicspath{{../}}

\begin{document}

\title{Lyapunov spectra and fluctuation relations: Insights from the Galerkin-truncated Burgers equation}

\author{Arunava Das}
%\email{}
\affiliation{Department of Physics, Indian Institute of Technology Kharagpur, Kharagpur - 721 302, India}%

\author{Pinaki Dutta}
%\email{}
\affiliation{Department of Physics, Indian Institute of Technology Kharagpur, Kharagpur - 721 302, India}%

\author{Kamal L. Panigrahi}
%\email{}
\affiliation{Department of Physics, Indian Institute of Technology Kharagpur, Kharagpur - 721 302, India}%

\author{Vishwanath Shukla}
\email{research.vishwanath@gmail.com}
\affiliation{Department of Physics, Indian Institute of Technology Kharagpur, Kharagpur - 721 302, India}%

\date{\today}
\begin{abstract}

The imposition of a global constraint of the conservation of total kinetic energy on a forced-dissipative Burgers equation yields a governing equation that is invariant under the time-reversal symmetry operation, $\{\mathcal{T}: t \to -t; u \to -u \}$, where $u$ is the velocity field. Moreover, the dissipation term gets strongly modified, as the viscosity is no longer a constant, but a fluctuating, state dependent quantity, which can even become negative in certain dynamical regimes. Despite these differences, the statistical properties of different dynamical regimes of the time-reversible Burgers equation and the standard forced-dssipative Burgers equation are equivalent, \`a la Gallavotti's  conjecture of \textit{equivalence of nonequilibrium ensembles}. We show that the negative viscosity events occur only in the thermalized regime described by the time-reversible equation. This quasi-equilibrium regime is examined by calculating the local Lyapunov spectra and fluctuation relations. A pairing symmetry among the spectra is observed, indicating that the dynamics is chaotic and has an attractor spanning the entire phase space of the system. The violations of the second law of thermodynamics are found to be in accordance with the fluctuation relations, namely the Gallavotti-Cohen relation based on the phase-space contraction rate and the Cohen-Searles fluctuation relation based on the energy production rate. It is also argued that these violations are associated with the effects of the Galerkin-truncation, the latter is responsible for the thermalization.

\end{abstract}

\maketitle

\section{Introduction}

The attempts to answer how irreversible macroscopic behaviour arises from fundamental reversible microscopic equations led to several debates in the past and continues to motivate new studies, but the general approach to answer this is now well accepted~\cite{lebowitz1993boltzmann}. Also, over the last few decades, fluctuation relations (FRs) have emerged as an important broad framework to study nonlinear, nonequilibrium processes. For example, FRs have been applied to a large class of systems driven far from equilibrium~\cite{evans1993probability,gallavotti1995dynamical,jarzynki1997freeenergy,kurchan1998fluctuation,Lebowitz1999,Chetrite2008,hayashi2010ft,carberry2004fluct,wang2002secondlawviol,evans2002fluctuation,annurev2008sevick}. For irreversible processes involving a large number of degrees of freedom, the forward time trajectories far outweigh the conjugate time-reversed trajectories. FRs enable this comparison by explicitly comparing the ratio of the probability of the forward-in-time trajectories to the probability of corresponding reversed trajectories, and show that the probability of the former is exponentially larger. As a consequence FRs can capture the statistics of entropy production~\cite{Merhav_2010,evans2005application}.

The one dimensional (1D) Burgers equation (BE), is a nonlinear hydrodynamic equation that is used to model a variety of problems in fluid dynamics~\cite{frisch2001burgulence,bec2007burgers,chowdhury2000statistical}. In this work, FRs are examined for the certain regimes of the Galerkin-truncated, forced-dissipative Burgers equation and its time-reversible formulation, introduced in~\cite{das2023rb}. Following Gallavotti's conjecture of \textit{equivalence of nonequilibrium ensembles}~\cite{gallavotti1996equivalence}, a time-reversible Burgers equation (RB) was derived by imposing a global constraint of the conservation of total kinetic energy~\cite{das2023rb}. One of the major consequences of such an imposition is that the transport coefficient becomes state dependent and fluctuates in time.

The conjecture has been examined for the three-dimensional (3D) Navier-Stokes equations (NSE)~\cite{biferale1998time,DePietro2018,Biferale_2018,2019phshuklaase,jaccod2021const,margazoglou2022nonequilibrium} and its two-dimensional (2D) version~\cite{Shukla2dequivalence2023}. In~\cite{das2023rb}, the conjecture was verified for the 1D BE and RB where different dynamical regimes of the Galerkin-truncated equations were studied. In these systems, one regime was identified to be dominated by truncation effects, exhibiting features that are consistent with the establishment of a quasi-equilibrium. In this work, we look at the statistics of various quantities in this regime and we find many instances of violation of the second law of thermodynamics. In particular, for the RB, the fluctuating transport coefficient can become negative at certain instances in the quasi-equilibrium regime. 

A variety of FRs are known to exist and they emerge as a consequence of the ensemble associated with a problem. For example, iso-energetic and iso-kinetic ensembles have different FRs~\cite{cohen1998note}. In order to characterize, the regimes of the BE and RB systems, two different FRs are considered here: The Gallavotti-Cohen fluctuation relation (GCFR)~\cite{gallavotti2020nonequilibrium} and the one introduced by Evans, Cohen and Searles~\cite{evans1993probability,evans2002fluctuation}. The former deals with phase-space compression rate, while the latter is concerned with a dissipation function that is related to the entropy production rate. In fact, these two FRs are not always equivalent depending on the ensemble involved~\cite{evans2005application}.

It is found that for the RB system, the instances of violations of the second law of thermodynamics are consistent with the above two FRs. Interestingly, Cohen-Searles FR is also satisfied for the Galerkin-truncated BE, when looking at the statistics of the energy injection rate or power. It is postulated that the appearance of such violations in the NSE and its variants are associated with the truncation effects, which arise due to the lack of sufficient length scales required for the direct numerical simulation of such a regime. Furthermore, the local Lyapunov spectra~\cite{GALLAVOTTI2004338,gallavotti2022navier,Gallavotti2021lyap,Gallavotti2020ensturbfr} of the BE and RB are examined. These provide a measure of the chaos inherent in the system, with large positive exponent being an indicator of an ergodic or a chaotic motion. A pairing symmetry in the exponents has been observed in the simulations of the 2D NSE and its time-reversible version in~\cite{GALLAVOTTI2004338}. Similar symmetry is observed in the BE and RB, albeit in a particular regime.

The remainder of this paper is organized as follows. In Sec. II, the BE and RB are introduced and an overview of their statistical properties is provided. In Sec. III, the expressions for the local Lyapunov exponents are presented and the appearance of a pairing symmetry is discussed. Section IV contains numerical validation of the FR for two non positive-definite quantities: the phase-space contraction rate and the energy injection rate. Section V presents the summary and the concluding remarks.

\begin{figure*}%[h!]
	\centering
		\includegraphics[width=0.45\linewidth]{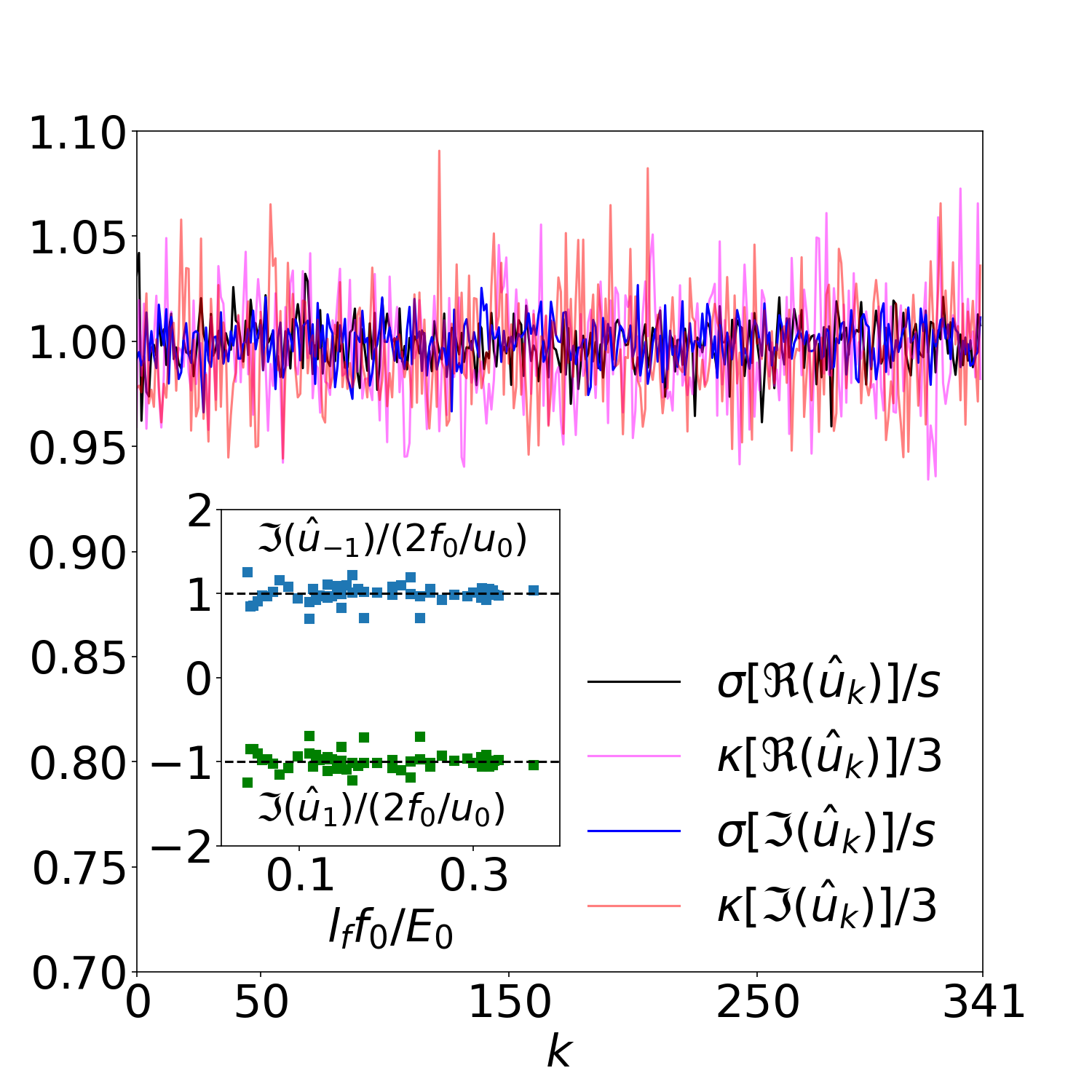}
		\put(-190,185){\bf{\large{(a)}}}
		\includegraphics[width=0.45\linewidth]{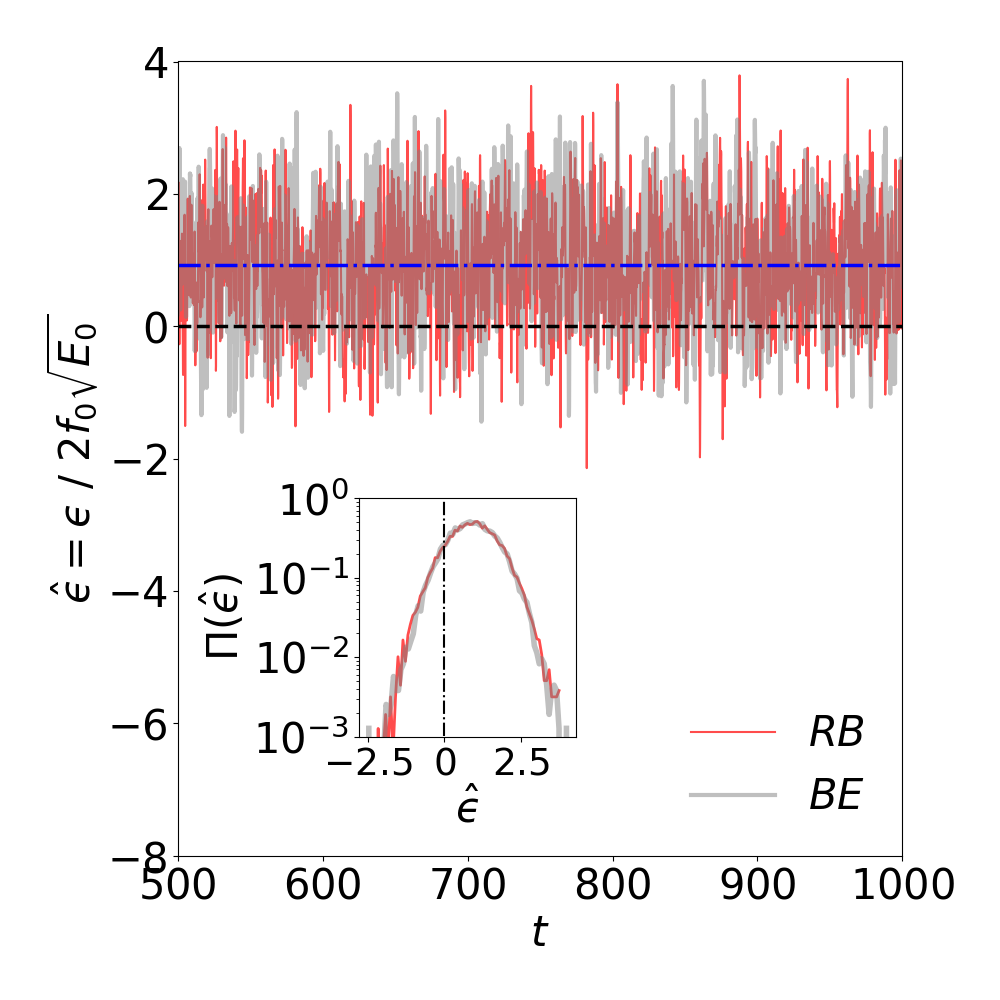}
		\put(-180,165){\bf{\large{(b)}}}
	\caption{ (a) Standard deviations and kurtosis of $\Re{ (\hat{u}_{k}) }$ and $\Im{ (\hat{u}_{k}) }$ are depicted, verifying that they are Gaussian with variance given by $s^2={u_{0}}^2/{8 \kmax}$. The inset shows the mean values of the Fourier modes at the forcing length scale for different values of $R=l_ff_0/E_0$ in the quasi-equilibrium regime. (b) Normalized time-series of the energy injection rate $ \epsilon(t)/2f_0\sqrt{E_0} $ of the RB and BE in the thermalized regimes for a simulation where both the systems are maintained at the same macroscopic energy ($R=0.28$). The blue dot-dashed line indicates the mean value of $\epsilon$. The black dashed line indicates the $\epsilon = 0$ value; the negative $\epsilon$ events are present in both the RB and the BE. The inset shows the PDFs of the energy injection rate for the two systems.}
	\label{quasieqilRBBE}
\end{figure*}

\section{The BE and RB}

The BE in the Fourier space is given by
\begin{equation}\label{eqn: BE}
	{\dot{\hat{u}}_{k}} = \sum_{k_{1}=-\infty}^{k_{1}=\infty}\frac{ik}{2} \hat{u}_{k_{1}} \hat{u}_{k-k_{1}} -\nu k^{2} \hat{u}_{k} +\hat{f}_{k},
\end{equation}
where $\hat{u}_{k}(t)$ is the velocity of the $k$'th Fourier mode of a real velocity field $u(x) = \sum_{k=-\infty}^{k=\infty} \hat{u}_{k}(t) \exp{ (ikx) }$, $\nu$ is the the viscosity and $\hat{f}_{k}$ is a local forcing term which is restricted to the $k_f = \pm 1$ forcing modes. The overdot represents derivative with respect to time. The chosen form of the forcing is given by
\begin{flalign*}
	\hat{f}_{\pm 1} & = \mp \frac{i}{2} f_{0},\\
	& = 0  \text{ } \quad \quad \forall \text{ } k \text{ } \neq \pm 1,
\end{flalign*}
which corresponds to $f(x) = f_0 \sin x$ in the real space, where $f_0$ is the forcing amplitude.
Also, $u(x) = u_{0} \sin x$ is used as initial conditions. This choice sets the zeroth Fourier mode $\hat{u}_{0} = 0$ and $\Re{(\hat{u}_{\pm1})}=0$, $\Im{(\hat{u}_{\pm1})}= \mp { u_{0}}/{2} $.

The RB is derived from the BE by imposing a thermostatting viscosity which restricts the energy of the system to a constant value, \textit{i.e}, $E_{0} = \sum_{k}|\hat{u}_{k}|^2/2$ is constant for all $t$. This is achieved by replacing $\nu$ with 
\begin{equation}
\nu_{r} = \frac{\epsilon}{\Omega}, 
\end{equation}
where $\epsilon = \sum_{k} \hat{f}_{k}^{*} \hat{u}_{k}$ is the energy injection rate and $\Omega = \sum_{k} k^2 |\hat{u}_{k}|^2$ is the strain rate squared term. As a result, the equation for the RB in the Fourier space is written as follows
\begin{equation}\label{eqn: RB}
	{\dot{\hat{u}}_{k}} = \sum_{k_{1}=-\infty}^{k_{1}=\infty}\frac{ik}{2} \hat{u}_{k_{1}} \hat{u}_{k-k_{1}} -\nu_{r} k^{2} \hat{u}_{k} + \hat{f}_{k}.
\end{equation}
Note that such a modification replaces the fluctuating energy in the BE with fluctuating viscosity in the RB. The RB now obeys the time-reversal symmetry $ \{ \mathcal{T} : t\rightarrow -t ;\hspace{2mm} u\rightarrow -u \} $, which is violated by the BE. 

The Fourier series representations of the BE and RB involve infinite terms. However, for the Galerkin-truncated equations, the Fourier coefficients of terms with $k$ larger than a particular wavenumber $\kmax$ are set to zero, i.e., $\hat{u}_k = 0~\forall~|k| \geq \kmax$. Hence, $\kmax$ is the largest Fourier mode that is retained. This procedure truncates all the previous sums such that they range from $-\kmax$ to $\kmax$, and leaves the BE and RB equations as a collection of finite number of coupled ODEs.

In Ref.~\cite{das2023rb}, the regimes of the RB were characterized by defining a parameter 
\begin{equation}\label{eq:R}
R = \frac{l_f f_0}{E_0}, 
\end{equation}
where $l_{f} f_{0}$ is the energy injected at the forcing scale, $l_{f}$, and $E_{0}$ is the total energy of the system (fixed by the initial conditions). Also, it was shown that the RB and BE satisfy the Gallavotti-Cohen equivalence conjecture in all the regimes~\cite{gallavotti1995dynamical,gallavotti1996equivalence,das2023rb}(details of the numerical simulations are provided in Appendix~\ref{app:dns}). This conjecture, modified for the BE and RB is stated as follows.
Let $BE_{\nu}^{\mathcal{K}}$ and $RB_{E_0}^{\mathcal{K}}$ be two ensembles corresponding to the BE and RB respectively. The states of the BE are characterized by the viscosity $\nu$, while those of the RB by the energy $E_0$. $\mathcal{K}$ denotes the largest wave-number $\kmax$ (or smallest length scales) of the systems. The equivalence conjecture then states that if $ {\langle E \rangle}_{BE_{\nu}^{\mathcal{K}}} = E_0 $, then for any local observable $O$ one has \[  \lim_{\substack{\mathcal{K} \rightarrow \infty \\ \nu \to 0}} {\langle O \rangle}_{BE_{\nu}^{\mathcal{K}}} = \lim_{\substack{\mathcal{K} \rightarrow \infty \\ \nu \to 0}} {\langle O \rangle}_{RB_{E_0}^{\mathcal{K}}}. \] The $\nu \to 0$ limit enters the right hand term of the above equation through the energy equivalence requirement, $ {\langle E \rangle}_{BE_{\nu}^{\mathcal{K}}} = E_0 $ (a local observable is one which depends on a finite number of $k$'s).

The conjecture as stated above is in one of its most stringent forms and it has been significantly weakened to include cases with finite non-vanishing viscosity and finite $k$~\cite{Gallavotti2018quant,gallavotti2020nonequilibrium,Gallavotti2021lyap,2019phshuklaase,margazoglou2022nonequilibrium,Shukla2dequivalence2023,das2023rb}.

In the RB, in regimes where the injected energy is large compared to the total energy for $R \gg 1$, a hydrodynamic shock dominated regime was observed, whereas for $R \ll 1$, a quasi-equilibrium state is seen. Moreover, the average $\langle \nu_r \rangle$ in the quasi-equilibrated regime was found to be very small, whereas $\langle \Omega \rangle$ was observed to be very large. The corresponding regimes in the BE were observed for comparable average energy, $E_0$, and the viscosity $\nu$ same as $\langle \nu_r \rangle$.

The quasi-equilibrium state is characterized by a presence of sufficiently random modes in the velocity field and a thermalized energy spectrum $E(k) = E_0/\kmax,$ with the energy equally distributed among the Fourier modes. The Gaussian nature of the Fourier modes is shown in Fig.~\ref{quasieqilRBBE}(a), where all the modes have independent, normally distributed real and imaginary parts. The mean is zero and the standard deviation is given by $s =  {u_{0}}/{\sqrt{8 \kmax}}$. Moreover, the kurtosis is very close to $3$, thereby, confirming the Gaussian nature of the velocity modes: $\Re(\hat{u}_{k}) \sim N(0, s^2)$ and $\Im(\hat{u}_{k}) \sim N(0, s^2)$.

The only modes which deviate from this are the imaginary components of the forcing modes at $ k_f = \pm 1 $. These components are still Gaussian, however, with a different mean. As such, $\Im{ (\hat{u}_{\pm1}) } \sim N( \mp 2f_0/u_0 |k_f| , s^2 )$. The means of $\Im{ (\hat{u}_{\pm1}) }$ are depicted in the inset of Fig.~1(a). Furthermore, the factor $2$ can be explained using the analytical solutions of the BE with sinusoidal forcing~\cite{das2023rb,banerjee2019fractional}, which in the inviscid limit is \[ u(x) = 2 \sgn{(x-\pi)} \sqrt{f_0} \sin(x/2). \] This form is due to the non-linear term causing a cascade from $k_f$ to the smallest scales; the largest scale in the solution is $k_f/2$.

The Gaussian distribution of the real and imaginary velocity Fourier components result in an energy spectrum given by $E(k) = |\hat{u}_{k}|^2 \sim \chi^2_2 $, \textit{i.e.}, $E(k)$ has a  Chi-squared distribution with mean and variance as $(2s^2,4s^4)$ for all the non-forcing modes. Furthermore, $\chi^2_2 \sim \Exp(1/2s^2)$. Hence, $E(k)$ follows an exponential distribution with parameter $1/2s^2$. This sets up the thermalized spectrum $E_0/\kmax$ that is observed in the one dimensional BE and RB.

Figure~\ref{quasieqilRBBE}(b) shows that the time-series of energy injection rate $\epsilon(t) = f_{0}\Im(\hat{u}_{-1})$, both for the RB and BE, exhibit negative events. The inset of Fig.~\ref{quasieqilRBBE}(b) shows the probability distribution function (PDF) of $\epsilon$, wherein it is clearly evident that the negative events constitute a significant fraction. These events are a consequence of the truncation effects modifying the distribution of the velocity at the forcing length scale. Note that the violations of the second law are not observed in the regimes other than the quasi-equilibrium one. In fact, it is argued that this lack of negative events is due to the attractor of the dynamics being of much smaller dimensions than that of the entire phase space in the said regimes.

For $\Omega = \sum_k k^2 | \hat{u}_k |^2$, it can easily be shown that the mean and variance are given by $2s^2 (\sum_k k^2) \approx 2/3E_0\kmax^2$ and $4s^4 (\sum_k k^4)$, respectively. Even though for such a linear combination of chi-squared/exponential random variables no closed form probability distribution exists, the large value of $\kmax$ implies that the central limit theorem is applicable and the distribution of $\Omega$ is effectively Gaussian. Surprisingly, $\nu_r = \epsilon/\Omega$ is also normally distributed. The mean and variance can be approximated such that $ \langle \nu_r \rangle \approx {\langle \epsilon \rangle} / {\langle \Omega \rangle} $ and $ \Var{ (\nu_r) } \approx \Var{ (\epsilon) } / {\langle \Omega \rangle}^2$.

\section{Local Lyapunov exponents and phase space contraction rate}

In this section, time-averaged local Lyapunov exponents are calculated from the Jacobian matrices of the BE and RB.  Once the Jacobian matrix is constructed from the concerned dynamical equation, the eigenvalues of the hermitian part of the Jacobian matrix give the local Lyapunov exponents at a particular time-instant. The analytical considerations and numerical calculations show that the depending on the regime, characterized by the value of $R$, both the systems could be either ordered  or highly chaotic. In fact, using expressions for the shock solution it is shown that in a regime free of truncation effects, there are no positive Lyapunov exponents and hence, little to no chaos. This is hardly surprising given that the BE is inherently devoid of chaos. However, in a heavily truncation dominated quasi-equilibrium regime, chaoticity arises and a pairing symmetry develops in the exponents. This is a clear indication that in the quasi-equilibrium regime, the attractor of the dynamics spans the entire phase space~\cite{gallavotti2020nonequilibrium}. Moreover, the expressions for phase-space contraction rates are determined for both the systems, which are later used to test fluctuation relations.

\subsection{Jacobian matrix for the BE and RB}
For the BE, the Jacobian matrix elements are given by
\begin{flalign*}
	J_{\alpha\beta} & = \frac{\partial \dot{{\hat{u}}}_{\alpha}}{\partial \hat{u}_{\beta}}\\
	& = i\alpha \hat{u}_{\alpha-\beta} - \nu \alpha^{2} \delta_{\alpha\beta}
	%& = \sum_{k_{1}}[\frac{ik}{2}\hat{u}_{k-k_{1}}\delta_{k_{1}h} + \frac{ik}{2} \hat{u}_{k_{1}}\delta_{(k-k_{1})h}]-\nu k^{2}\delta_{kh} \\
	%& = \frac{ik}{2}\hat{u}_{k-h} + \frac{ik}{2}\hat{u}_{k-h} - \nu k^{2}\delta_{kh}.\\
\end{flalign*}
The hermitian part is then constructed as
\begin{equation}\label{eq:BEHemitian}
	H_{\alpha\beta} = \frac{1}{2}(J_{\alpha\beta}+J_{\beta\alpha}^{*}) = \frac{i}{2}(\alpha-\beta)\hat{u}_{\alpha-\beta} - \nu \alpha^2\delta_{\alpha\beta},
\end{equation}
where it is assumed that $-\kmax\leq \alpha,\beta\leq \kmax$. 

Following the same procedure as above, the Jacobian matrix elements, $J^{R}_{\alpha\beta}$, for the RB are given by
\begin{multline*}
	J^{R}_{\alpha\beta}  =
	i\alpha \hat{u}_{\alpha-\beta}-\frac{\alpha^{2}}{\Omega}
	\hat{u}_{\alpha}\hat{f}_{-\beta} +
	\nu_{r} \big( \frac{\alpha^2 \beta^2}{\Omega}\hat{u}_{\alpha}\hat{u}_{-\beta} - \alpha^{2} \delta_{\alpha\beta} \big).
\end{multline*}
The hermitian part, $H_{\alpha\beta}^{R}$ = $ \big[ J_{\alpha\beta}^{R}+(J_{\beta\alpha}^{R})^* \big]/2 $, becomes
\begin{multline}\label{eq:RBHermitianJac}
	H_{\alpha\beta}^{R} = \frac{i}{2}(\alpha-\beta)\hat{u}_{\alpha-\beta}-\frac{1}{2\Omega}(\alpha^{2}\hat{u}_{\alpha}\hat{f}_{-\beta}+\beta^2 \hat{u}_{-\beta}\hat{f}_{\alpha})\\
	+ \nu_{r}(\frac{\alpha^2 \beta^2}{\Omega}\hat{u}_{\alpha}\hat{u}_{-\beta} - \alpha^{2} \delta_{\alpha\beta}).
\end{multline}

\begin{figure*}%[h!]
	\centering
		\includegraphics[width=0.45\linewidth]{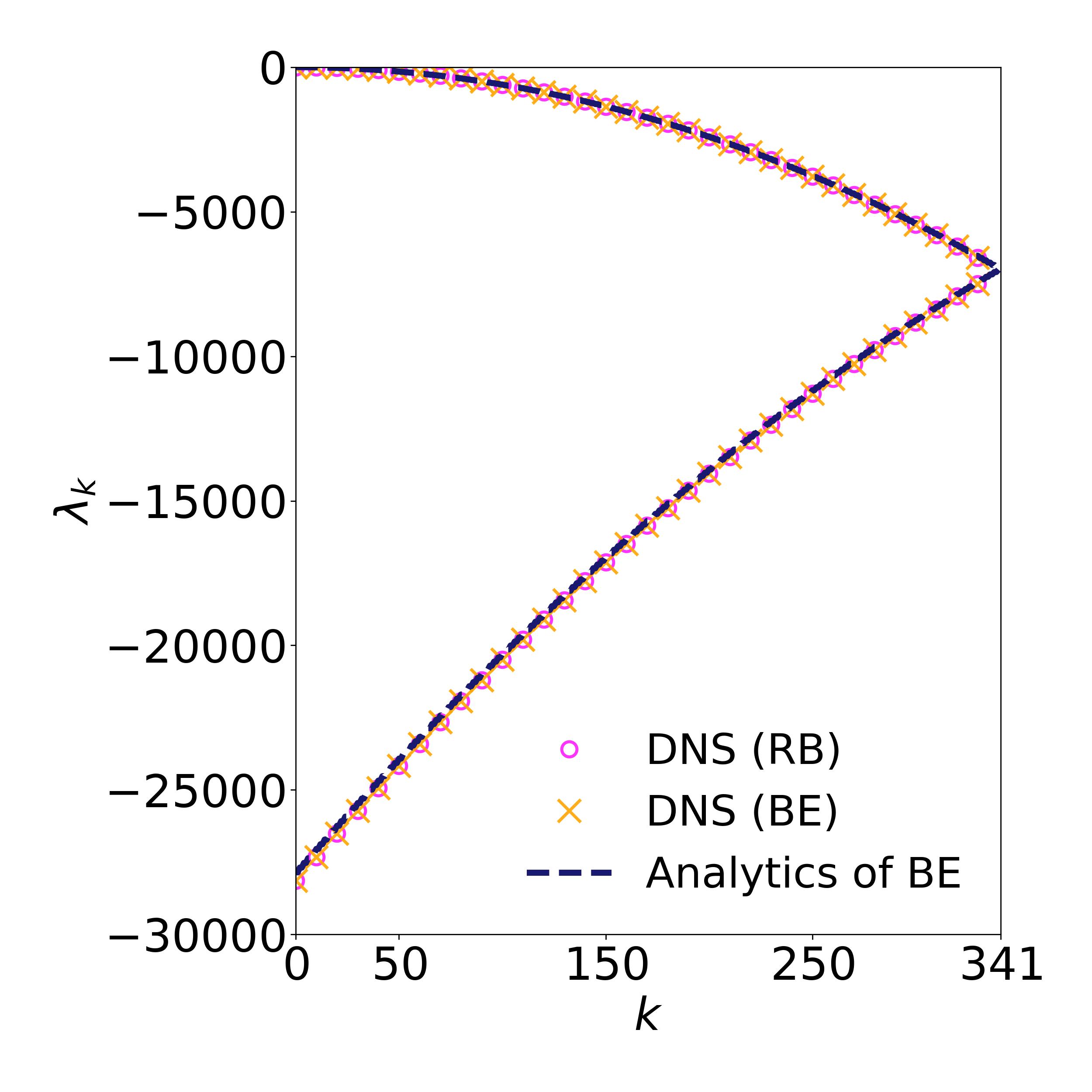}
		\put(-160,195){\bf{\large{(a)}}}
		\includegraphics[width=0.45\linewidth]{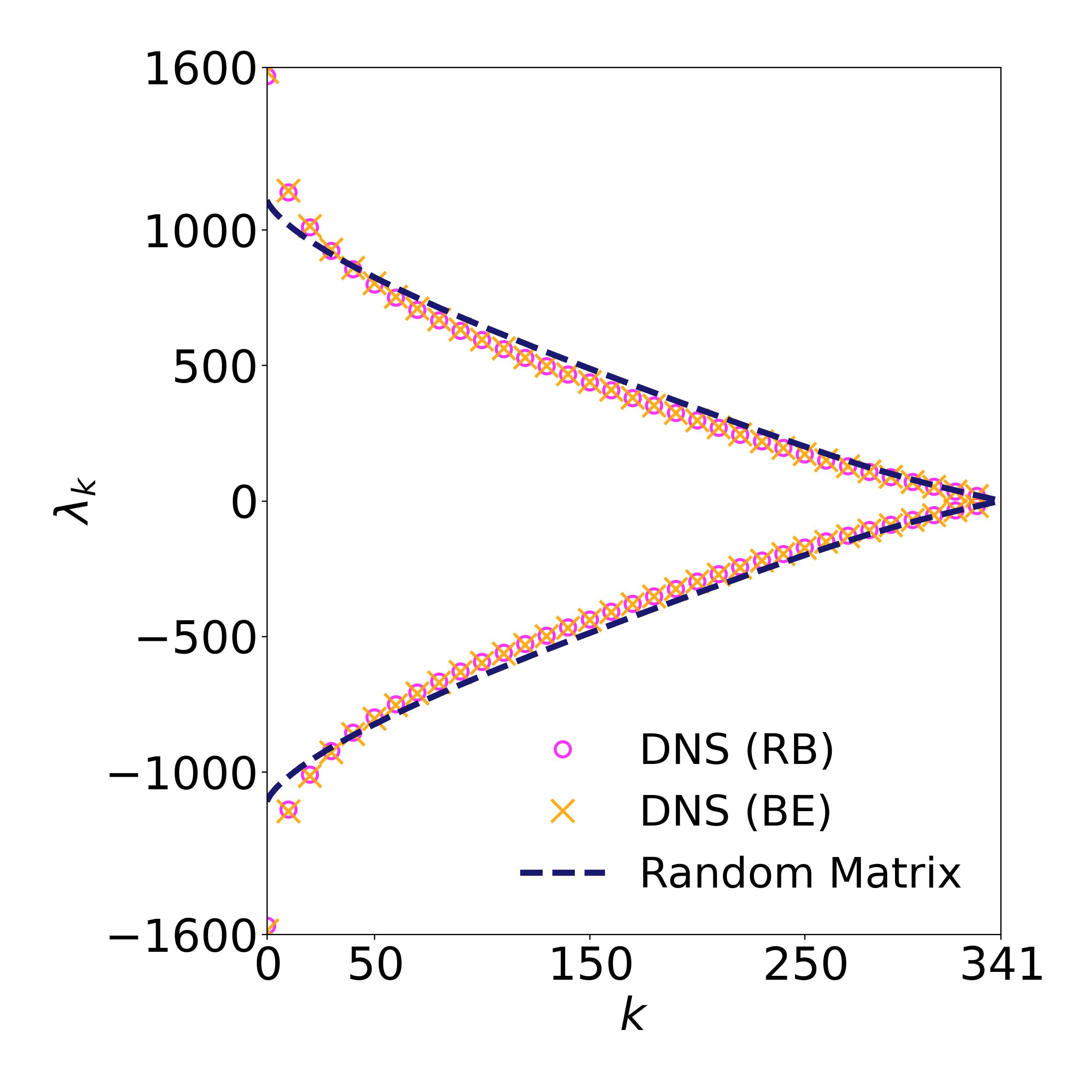}
		\put(-160,195){\bf{\large{(b)}}}
	\caption{(a) Local Lyapunov spectra for the RB (magenta `$\circ$' markers) and BE (yellow `$\times$' markers) in the shock dominated regime. The thick blue color dashed line gives the Lyapunov spectra for the matrix given by Eq.~\eqref{matrixBE} with $\hat{u}_{\alpha}$ = ${2}/{i\alpha t}$. The results from the DNSs of the BE, RB and the analytical calculations show a very good agreement. (b) Local Lyapunov spectra corresponding to the thermalized regime. An excellent agreement between the RB and the BE is seen. The thick blue color dashed line depicts the Lyapunov spectra for a Gaussian distributed random matrix. A pairing symmetry is evident in this regime. }
	\label{fig:lypspectra}
\end{figure*}

\subsection{Local Lyapunov spectra}

The local Lyapunov spectra for the RB and BE depend on the nature of the Fourier coefficients of the solutions, \textit{i.e.}, whether the system is in a shock dominated, but truncation effects free state, ($R\gg 1$), or the quasi-equilibrium regime, ($R \ll 1$). In the presence of shocks, $\hat{u}_k \sim k^{-1}$. In fact, it can be shown that the most dominant component $\hat{u}_k$ is given by $\hat{u}_k = {2}/{ikt}$. For the quasi-equilibrium regime, all the Fourier modes $\hat{u}_k$ are normally distributed. However, in order to simplify the calculation and to obtain a quick estimate in this regime, the Fourier modes are set to a constant value, $\hat{u}_k = c$. This is also used to highlight the features that are responsible for the pairing symmetry. Later on, the random nature of $\hat{u}_k$ is restored. 

For the BE, the Hermitian part of the Jacobian matrix, Eq.~\eqref{eq:BEHemitian}, can be expressed as
\begin{equation}\label{matrixBE}
	H_{\alpha\beta}  = A - \nu D,
\end{equation}
where $A_{\alpha \beta} = \frac{i}{2}(\alpha-\beta)\hat{u}_{\alpha-\beta}$ and $D_{\alpha \beta} = \alpha^2\delta_{\alpha\beta}$. Next, the following two cases are considered.

\textit{(i)} For $\hat{u}_{\alpha} = {2}/{i\alpha t}$,
\begin{equation}\label{shocklya}
	A_{\alpha \beta} = \begin{cases}
		{1}/{t}, & \mbox{$|\alpha-\beta|\leq \kmax$}\\
		0, & \text{otherwise}.
	\end{cases}\
\end{equation}

\textit{(ii)} For $\hat{u}_{\alpha} = c$ (a constant),
\begin{equation*}
	A_{\alpha \beta} = \frac{ic}{2}(\alpha-\beta) \text{ } \forall \text{ } |\alpha-\beta| \leq \kmax.
\end{equation*}
Taking $\kmax = 2$, for the case \textit{(i)}, $A$ becomes a $[5 \times 5]$ matrix with the elements:
\begin{equation}
 A_{\alpha\beta}= \frac{1}{t}\begin{bmatrix}
	0 & 1 & 1 & 0 & 0\\
	1 & 0 & 1 & 1 & 0\\
	1 & 1 & 0 & 1 & 1\\
	0 & 1 & 1 & 0 & 1\\
	0 & 0 & 1 & 1 & 0
\end{bmatrix}.
\end{equation}

Averaged over the time interval $\tau$, Eq.~\eqref{matrixBE} becomes
\begin{equation}
\langle H \rangle_t = \frac{1}{\tau} log(\frac{\tau}{\delta}) A ^ {\prime} - \nu D,
\end{equation}
where $\langle \rangle_t$ denotes time averaging and $\delta > 0$ is a small number and $A^{\prime}$ is the constant matrix above. The eigenvalues of $H$ at $t = 1$ and $\nu = 1$ are: $\lambda_{1}\approx -4.51882$, $\lambda_{2}\approx -4.41421$, $\lambda_{3}\approx 1.90932$, $\lambda_{4}\approx-1.58579$, and $\lambda_{5}\approx 1.3907$. The negative contribution to the eigenvalues come from the matrix $D$, and as a result for large $\tau$, the positive eigenvalues disappear. This is a clear indication of the lack of chaotic dynamics in the shock dominated BE regimes.

The matrix, $A_{\alpha\beta}$, for the case \textit{(ii)} at $\kmax=2$ is given by
$$ A_{\alpha\beta}= \frac{ic}{2}\begin{bmatrix}
	0 & 1 & 2 & 0 & 0\\
	-1 & 0 & 1 & 2 & 0\\
	-2 & -1 & 0 & 1 & 2\\
	0 & -2 & -1 & 0 & 1\\
	0 & 0 & -2 & -1 & 0
\end{bmatrix}$$

In the expression, $\langle H \rangle_{t} = A - \nu D$, without loss of generality $c$ is fixed at $1$ in the quasi-equilibrium regime. Therefore, in the limit,  $\nu \rightarrow 0$, the eigenvalues are $0, \pm 1.07635, \pm 0.692322$. Thus, a paring symmetry is observed in the eigenvalues, which was absent in case \textit{(i)}. In fact, the matrices $A_{\alpha\beta}$ at any given value of $\kmax$ are skew-symmetric for $\nu \rightarrow 0$, and hence its eigenvalues have a pairing symmetry.

For the RB, the Hermitian part of the Jacobian matrix, Eq.~\eqref{eq:RBHermitianJac}, can be expressed as
\begin{equation}
	H = A - \nu_{r} D - \frac{F}{\Omega} + \frac{\nu_{r} G}{\Omega},
\end{equation}
where $F_{\alpha \beta}$ = $(\alpha^{2}\hat{u}_{\alpha}\hat{f}_{-\beta}+\beta^{2}\hat{u}_{-\beta}\hat{f}_{\alpha})/2$ and $G_{\alpha\beta}$ = $\alpha^{2}\beta^{2}\hat{u}_{\alpha}\hat{u}_{-\beta}$. The matrices $A$ and $D$ are same as the ones that appear in the BE system. Once again, the following two cases are considered.

\textit{(i)} For $\hat{u}_{\alpha} = 2/{i\alpha t}$, $F_{\alpha \beta} = (\alpha \hat{f}_{-\beta} - \beta \hat{f}_{\alpha})/ it$. The forcing is localized at the large length scale, given by $\alpha = 1$. Thus, the only non-zero components of $F$ are $F_{1,1} = 1/t$, $F_{1,-1} = -1/t$ and $F_{\alpha \beta} = 0$ $\forall$ $\alpha,\beta \neq \pm 1$. Hence, $F_{\alpha\beta}$ for $\kmax=2$ is given by
\begin{equation}
F_{\alpha\beta}= \frac{1}{t}\begin{bmatrix}
	0 & 0 & 0 & 0 & 0\\
	0 & 1 & 0 & -1 & 0\\
	0 & 0 & 0 & 0 & 0\\
	0 & -1 & 0 & 1 & 0\\
	0 & 0 & 0 & 0 & 0
\end{bmatrix}.
\end{equation}
Note that for any value of $\kmax$, but with $\hat{f}_k$ restricted to a single mode (at $k = 1$), $F$ is a matrix of rank $1$ having at most one non-zero eigenvalue. This does not contribute to pairing. Similarly, $G_{\alpha \beta}$ = ${4}\alpha \beta/{t^{2}}$ has at most one non-zero eigenvalue. Therefore, the Hermitian part of the Jacobian matrix can be expressed as
\begin{equation*}
	H = \big[A^{\prime}/t - \nu_{r}(t)D \big] - \frac{1/t}{\Omega(t)}F^{\prime} + \frac{\nu_{r}(t)}{\Omega(t)}{4}G^{\prime}/{t^{2}},
\end{equation*}
where $G_{\alpha\beta}^{\prime}$ = $\alpha \beta$, $F^{\prime}$ = $(\alpha \hat{f}_{-\beta} - \beta \hat{f}_{\alpha})$, and $A^{\prime}$ is the same as in the BE. Next, $H$ is averaged over a time interval $\tau$
\begin{equation}
	\langle H \rangle = \big\langle A^{\prime}/t \big\rangle - \big\langle \nu_{r}(t) \big\rangle D - \big\langle \frac{1/t}{\Omega(t)} \big\rangle F^{\prime} + \big\langle \frac{4\nu_{r}(t)}{t^2\Omega(t)} \big\rangle  G^{\prime}.
\end{equation}

For large $t$, the last two terms in the above equation become sufficiently small and a behaviour approximately similar to the BE case \textit{(i)} is recovered
\begin{equation}
 \langle H \rangle \approx \frac{1}{\tau} log(\frac{\tau}{\delta})A^{\prime} - \big\langle \nu_{r}(t) \big\rangle D.
\end{equation}
Hence, positive eigenvalues are absent at large times.

\textit{(ii)} The choice of $\hat{u}_{\alpha} = c$ leads to $F_{\alpha\beta}$ = ${c}(\alpha^{2} \hat{f}_{-\beta} + \beta^{2} \hat{f}_{\alpha})/2$. Therefore, $F_{1,1} = 0$, $F_{1,-1} = ic/{2}$ and $F_{\alpha \beta} = 0$ for any other combination of $\alpha$ and $\beta$. For $\kmax = 2$, this can be expressed as
\begin{equation}
 F_{\alpha\beta}= \frac{ic}{2}\begin{bmatrix}
	0 & 0 & 0 & 0 & 0\\
	0 & 0 & 0 & -1 & 0\\
	0 & 0 & 0 & 0 & 0\\
	0 & 1 & 0 & 0 & 0\\
	0 & 0 & 0 & 0 & 0
\end{bmatrix}.
\end{equation}
$F$ is of rank $2$; thus, the pairing symmetry is possible and it holds for all the values of $\kmax$. Also, $G_{\alpha \beta}$ = $c^{2}\alpha \beta$. Note that $G$ matrix has at most one non-zero eigenvalue; hence, it does not contribute to pairing symmetry. Therefore, the Hermitian part of the Jacobian matrix for this case can be written as
\begin{equation*}
	H = [A - \nu_{r}(t)D] - \frac{i c}{2 \Omega(t)}F^{\prime} + \frac{\nu_{r}(t) c^2}{\Omega(t)}G^{\prime},
\end{equation*}
where $G^{\prime}_{\alpha \beta}$ = $\alpha^2 \beta^2$ and $F^{\prime}_{\alpha \beta}$ = $\alpha^{2} \hat{f}_{-\beta} + \beta^{2} \hat{f}_{\alpha}$.
The time-averaged $H$ is given by
\begin{equation*}
	\langle H \rangle = A - \langle \nu_{r}(t)\rangle D - \frac{i c F^{\prime}}{2} \langle 1/\Omega(t)\rangle + c^2\langle \frac{\nu_{r}}{\Omega} \rangle G^{\prime}
\end{equation*}
In the quasi-equilibrium regime, $\langle \Omega \rangle$ is very large and the viscosity $\langle \nu_{r} \rangle \to 0$. Thus, $\langle H \rangle \approx A$; which implies that the pairing symmetry is possible.

The local Lyapunov spectra for the shock dominated regime (free of truncation effects) and the quasi-equilibrium regime, are shown in Fig.~\ref{fig:lypspectra} (a) and (b), respectively. Fig.~\ref{fig:lypspectra} (a) shows that the Lyapunov spectra calculated analytically agrees perfectly with the numerically determined spectra.

Recall that so far for the analytical determination of the Lyapunov spectra in the quasi-equilibrium regime of the BE and RB, the velocity Fourier modes were assumed to be constant. However, these Fourier modes, $\hat{u}_{\alpha}$, fluctuate in time and are normally distributed. To account for this latter feature, a hermitian random matrix (see Eq.~\eqref{matrixBE}) is constructed by drawing $\hat{u}_{\alpha}$ from a Gaussian distribution $\sim N(0,s^2)$, except for the matrix elements with indices $\alpha = \pm |k_f|$ (forcing modes). These forced modes are chosen from the Gaussian distribution~$ N( \mp 2f_0/u_0 |k_f| , s^2 )$.

The local Lyapunov spectra for the quasi-equilibrium regime are shown in Fig.~\ref{fig:lypspectra} (b), wherein the numerically determined spectra for the BE and RB are plotted using the yellow crosses (`$\times$')  and the magenta circles (`$\circ$') , respectively. These overlap with each other perfectly. The Lyapunov spectra obtained using the random matrix model, depicted using blue dashed-line, is good in agreement with the numerically determined ones.

The development of the pairing symmetry in the Lyapunov exponents, in the thermal regime of the BE and RB, is unequivocally evident both in numerical simulations, as well as in the above simplified analytical treatments. Such a pairing symmetry of Lyapunov exponents is a well known feature of Hamiltonian systems (see footnote 18 of~\cite{gallavotti2020nonequilibrium}). It has further been extended to systems with a friction force in~\cite{lyapairfrictionDressler1998} and to cases with an isokinetic thermostat where the transport coefficient corresponding to the frictional force is replaced by a state dependent one. The presence of pairing symmetry shows that the attractor of the dynamics spans the entire phase space; this expansion of the attractor to the entire phase space can lead to events in violation of the second law of thermodynamics. The effect of the pairing symmetry on the attractor of the dynamics and subsequently its time-reversible nature will be discussed further in the concluding section.

\subsection{Phase space contraction rate}

The phase space contraction rate, $\Lambda$, can be obtained from the Jacobian matrix as
\begin{equation}
\Lambda(t) = - \Tr{ \big[ J \big] }.
\end{equation}
For the BE, it is
\begin{equation}\label{eq:pspBE}
	\Lambda = \nu \sum_{k} k^2 = \nu \big[ \kmax(\kmax+1)(2\kmax+1)/3 \big],
\end{equation}
which is clearly a constant of motion. For the RB, the phase space contraction rate is a time-fluctuating quantity and is given by
\begin{equation}\label{eq:pspRB}
	\Lambda(t) = \nu_r \sum_{k} k^2 + \frac{1}{\Omega} \sum_{k} k^2 (\hat{f}_{k}^{*}\hat{u}_{k}) - \frac{\nu_r}{\Omega} \sum_{k} k^4 | \hat{u}_{k} |^2 .
\end{equation}

\section{Fluctuation Relations}
\label{sec:fr}

\begin{figure}
	\includegraphics[width=0.8\linewidth]{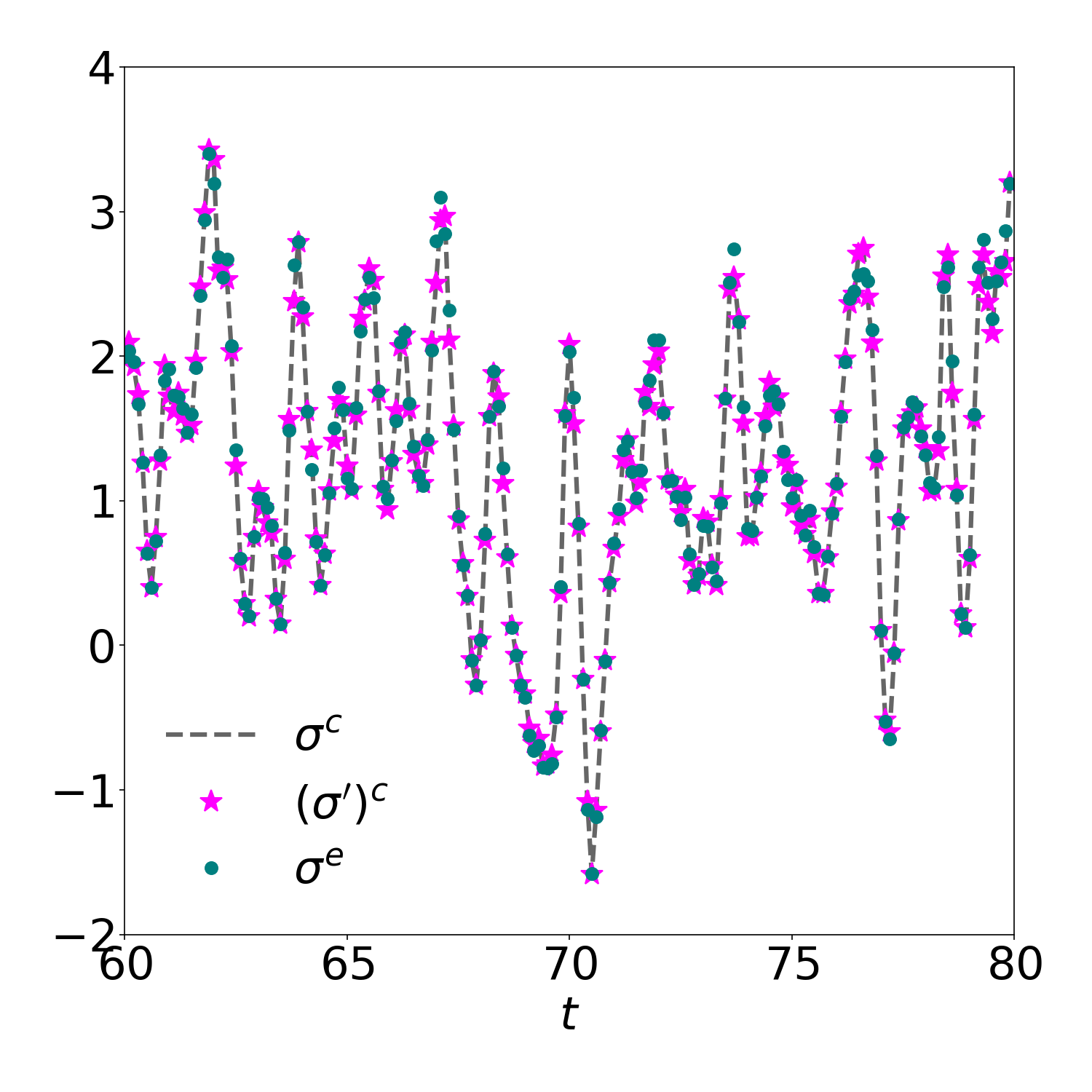}
	\caption{Comparison of the phase space contraction rate, $\sigma^c$ (black dashed line), analogous quantity ${\sigma^{\prime}}^c$ (magenta $\star$ symbols, see Eq.~\eqref{eq:analogpsc}) and the entropy production rate $\sigma^e$ (green filled circles) suggests the equivalence of these quantities.
	}
	\label{fig:contraction}
\end{figure}

\begin{figure*}%[h!]
	\centering
		\includegraphics[width=0.45\linewidth]{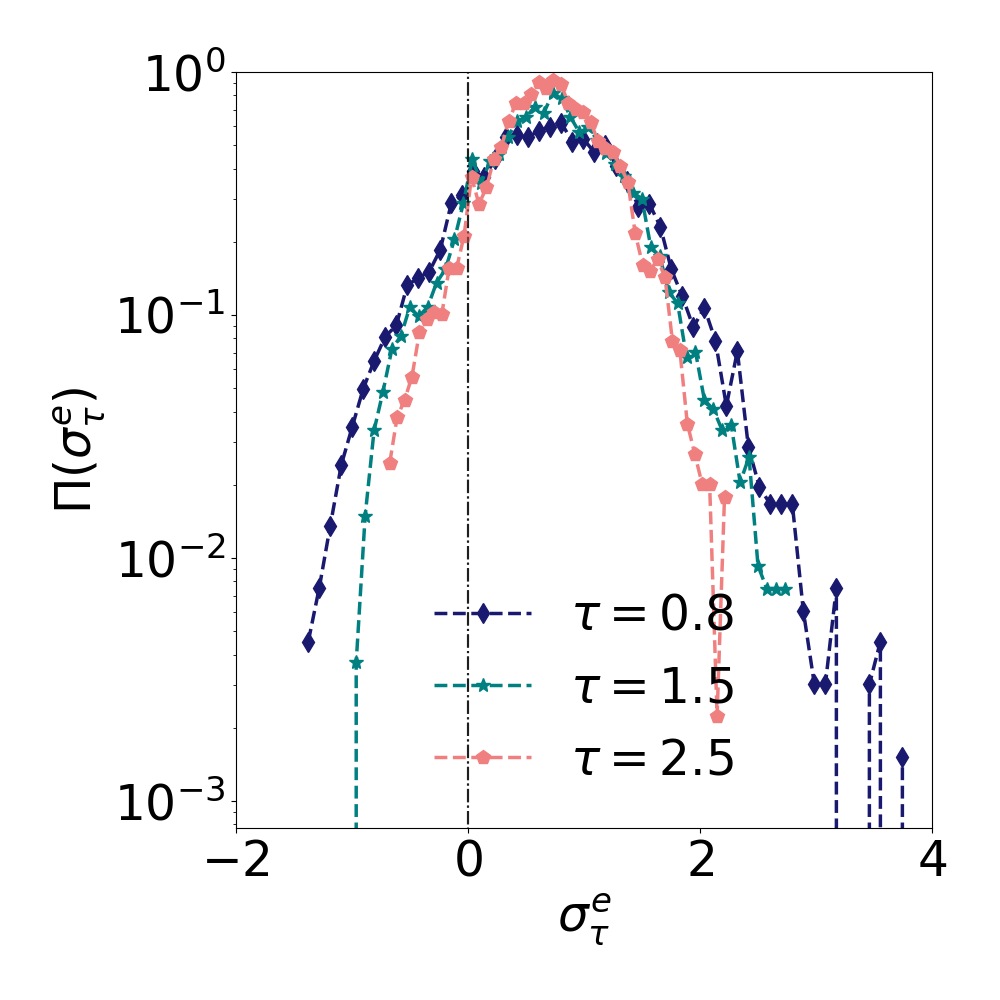}
		\put(-175,195){\bf{\large{(a)}}}
		\includegraphics[width=0.45\linewidth]{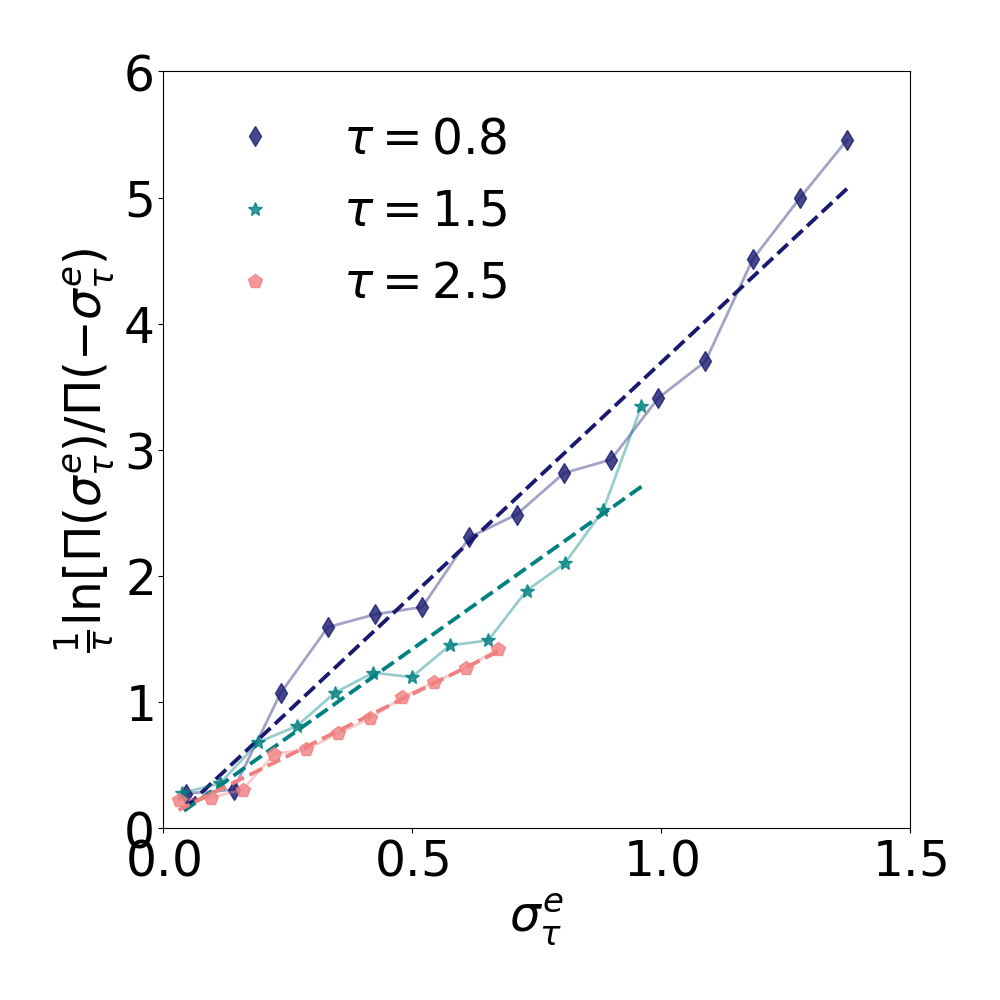}
		\put(-45,50){\bf{\large{(b)}}}\\
		\includegraphics[width=0.45\linewidth]{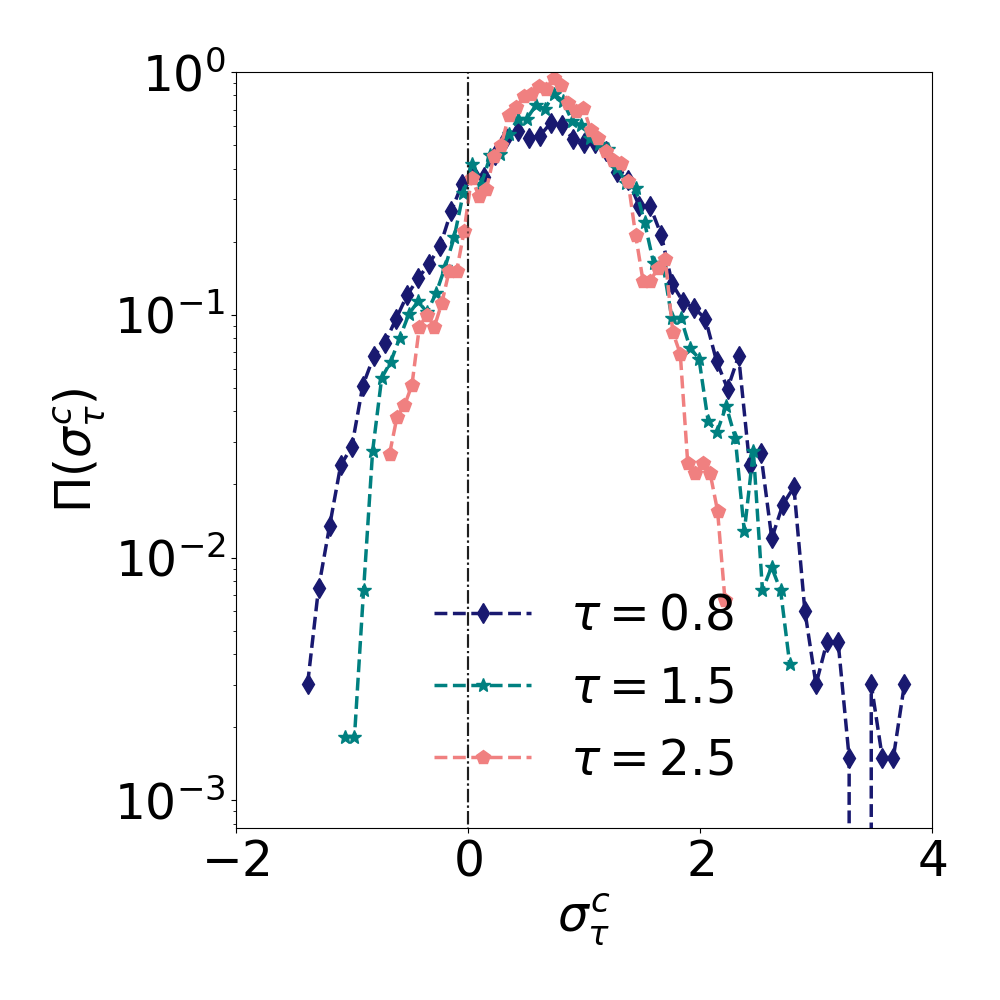}
		\put(-175,195){\bf{\large{(c)}}}
		\includegraphics[width=0.45\linewidth]{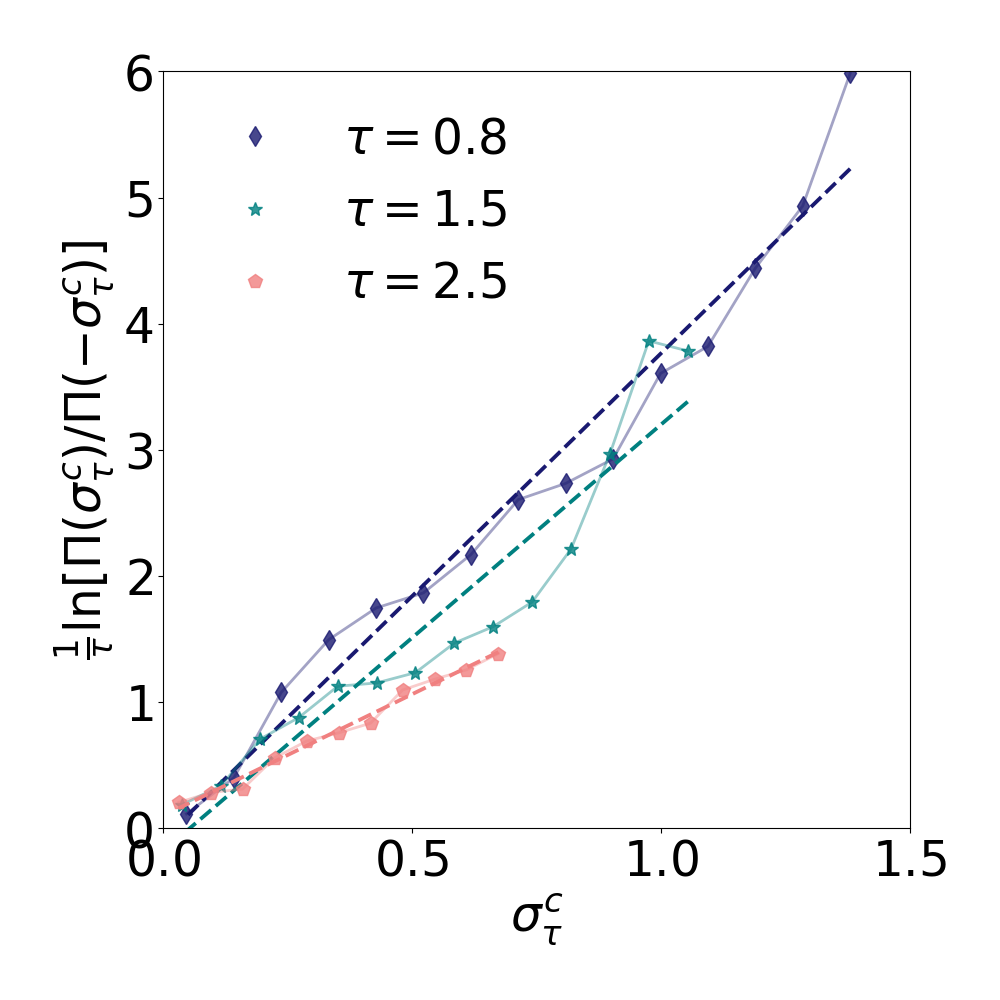}
		\put(-45,50){\bf{\large{(d)}}}
	\caption{Fluctuation relations for the RB. (a) PDFs of the entropy production rate, $\sigma^e_{\tau}$, averaged over the interval $\tau$; (b) plots of $(1/\tau) \Pi(\sigma^e_{\tau})/\Pi(-\sigma^e_{\tau})$ vs $\sigma^e_{\tau}$. (c) PDFs of the phase space contraction rate, $\sigma^c_{\tau}$, averaged over the interval $\tau$; (d) plots of $(1/\tau) \Pi(\sigma^c_{\tau})/\Pi(-\sigma^c_{\tau})$ vs $\sigma^c_{\tau}$. The linear behavior of curves in (b) and (d) suggests that the FRs hold valid. Here, $R=0.17$ and the dashed straight lines are the best-fit obtained using the method of least-squares.
}
	\label{fig:FR}
\end{figure*}

\begin{figure}
\includegraphics[width=0.8\linewidth]{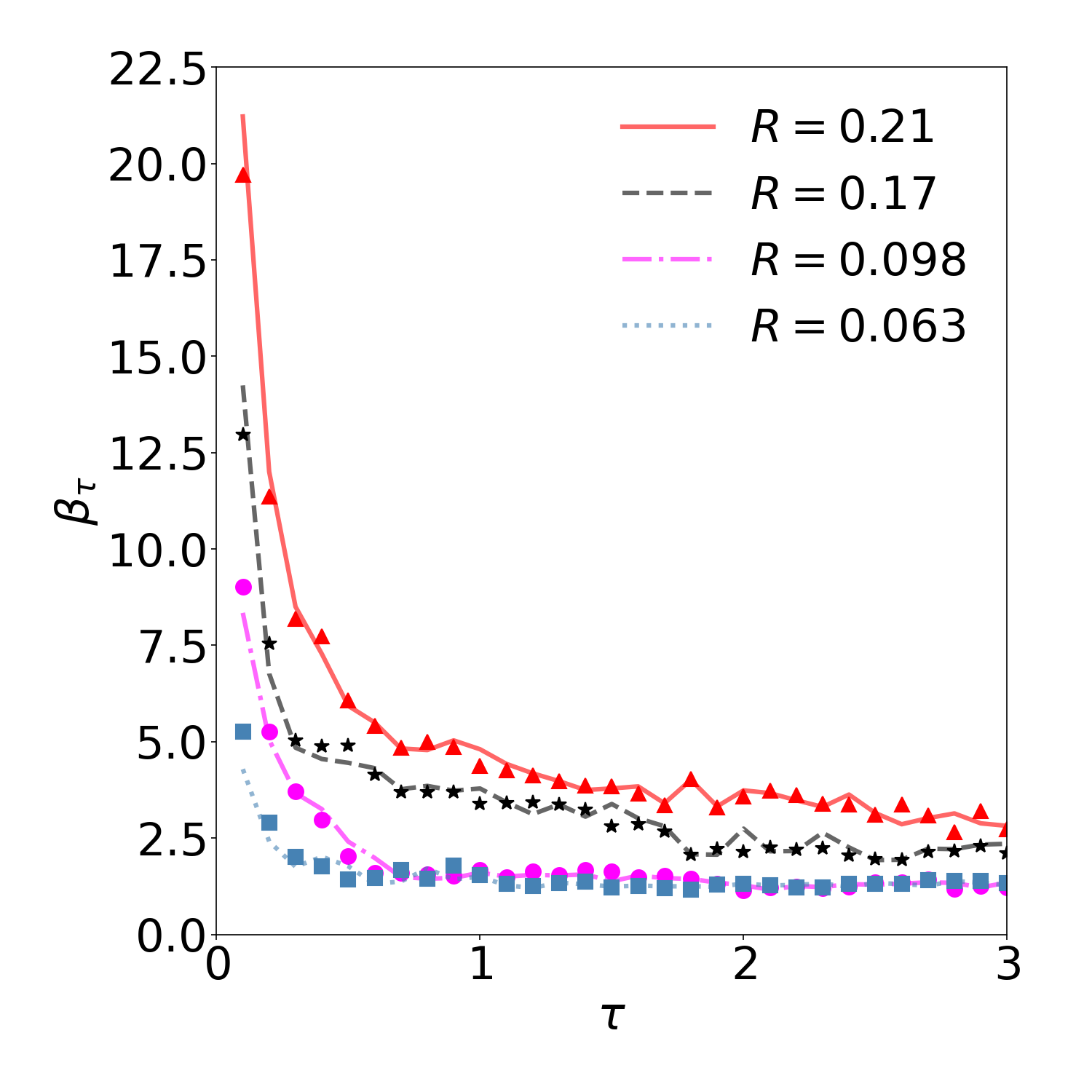}
\caption{Convergence of $\beta$. The values of $\beta$ in the thermalized regime converge to a constant value, as $\tau$ is increased. The lines depict the values of $\beta$ for the phase space contraction rate$\sigma^c$, while the markers indicate the same for the entropy production rate $\sigma^e$ for a given $R$. The good agreement between the two suggests that both the quantities are equivalent.
}
\label{fig:betaconverge}
\end{figure}

\begin{figure*}%[h!]
	\centering
		\includegraphics[width=0.45\linewidth]{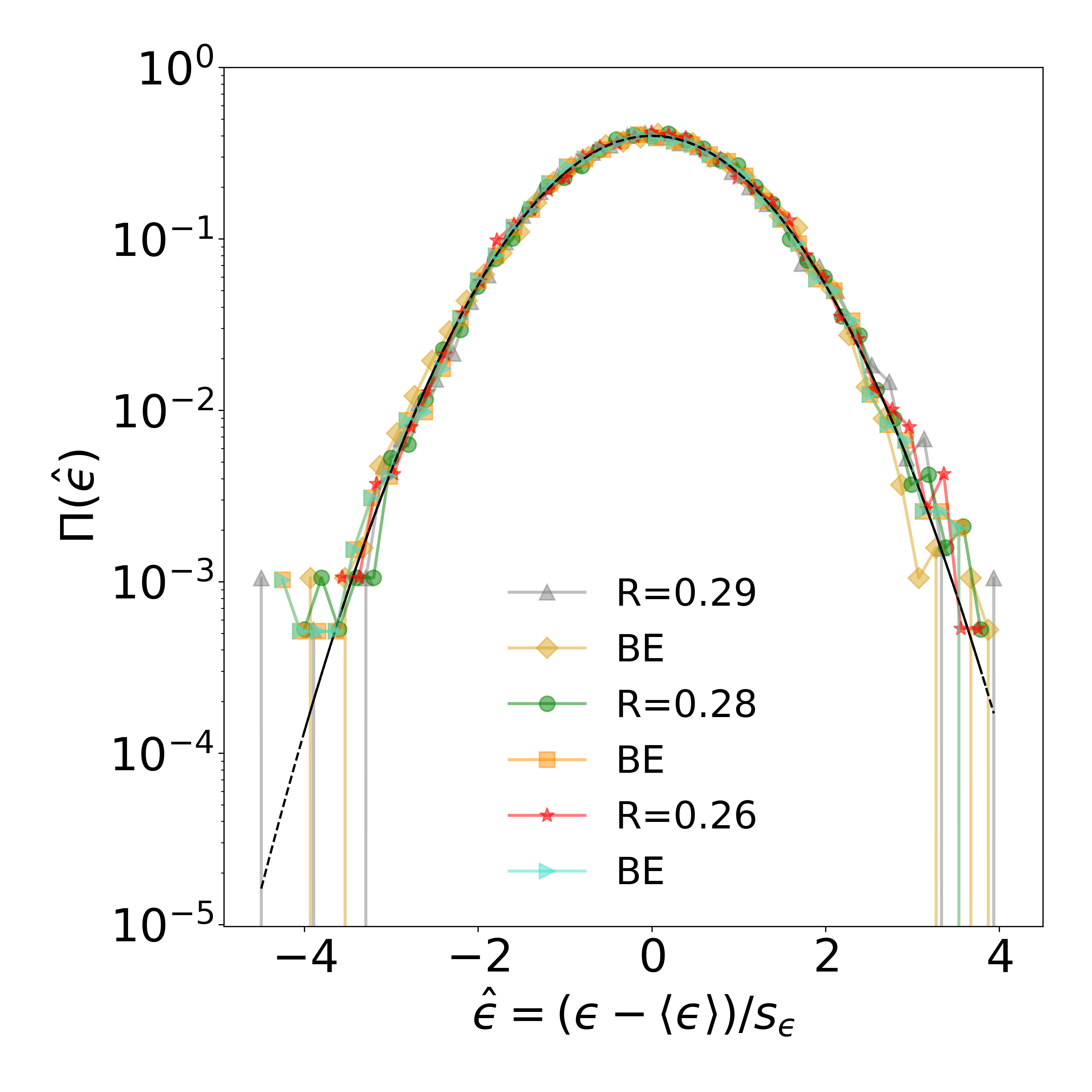}
		\put(-175,195){\bf{\large{(a)}}}
		\includegraphics[width=0.45\linewidth]{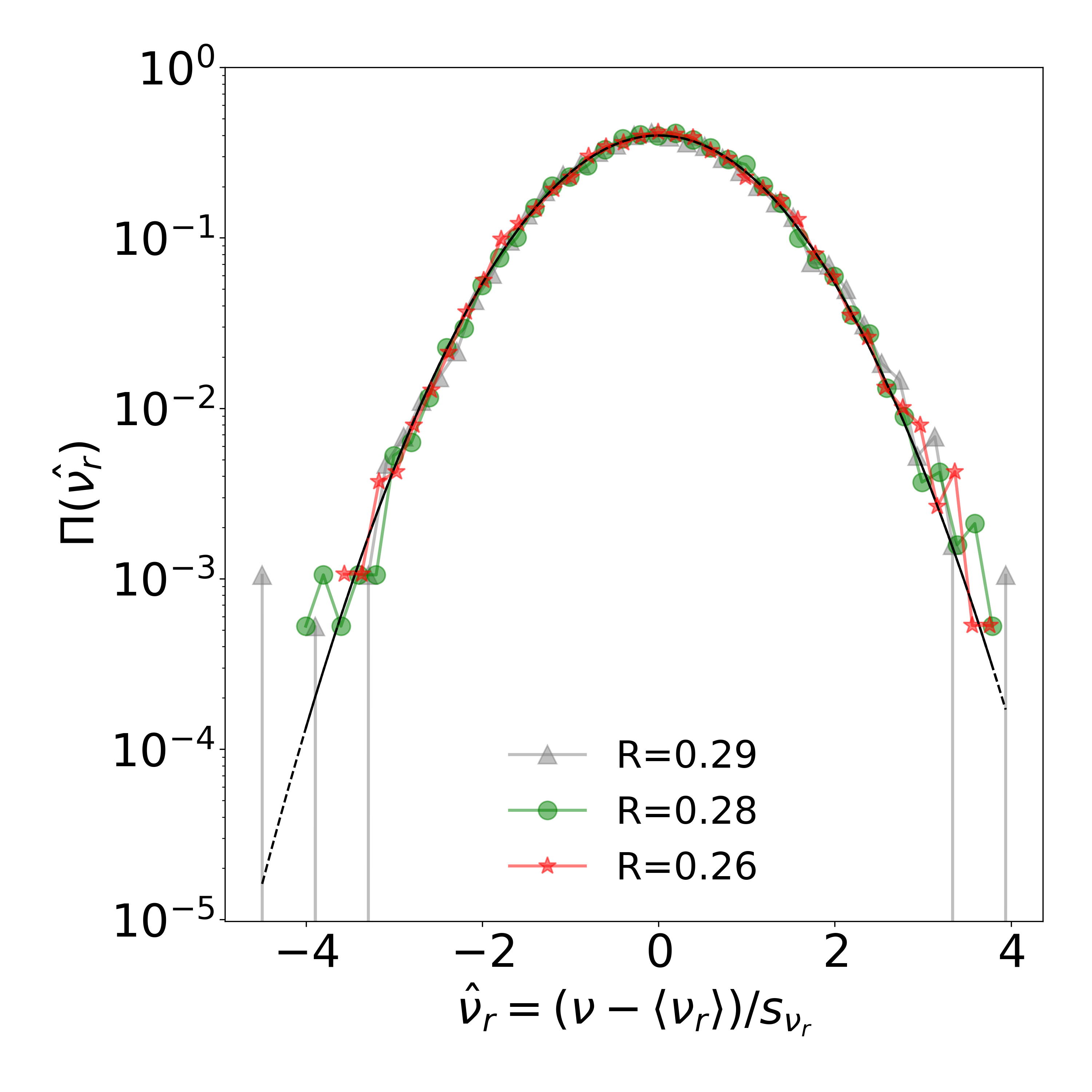}
		\put(-175,195){\bf{\large{(b)}}}\\
		\includegraphics[width=0.45\linewidth]{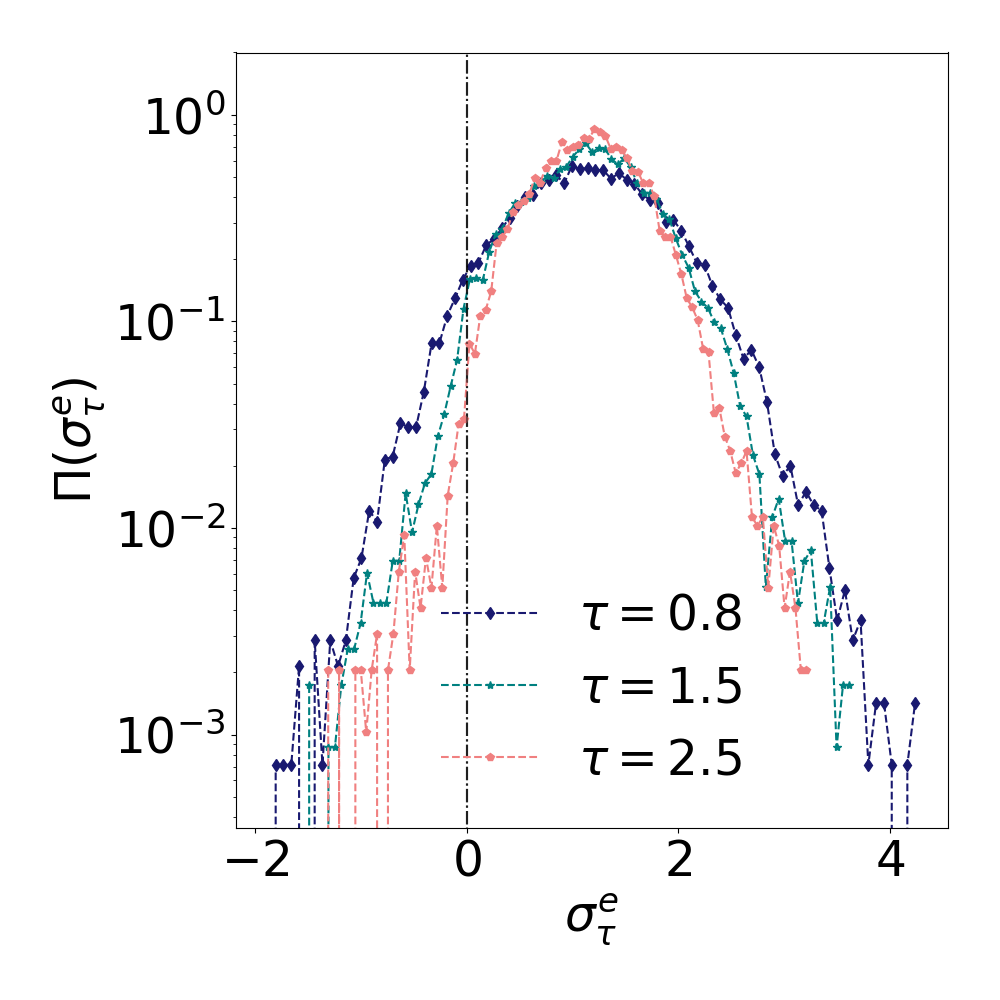}
		\put(-175,195){\bf{\large{(c)}}}
		\includegraphics[width=0.45\linewidth]{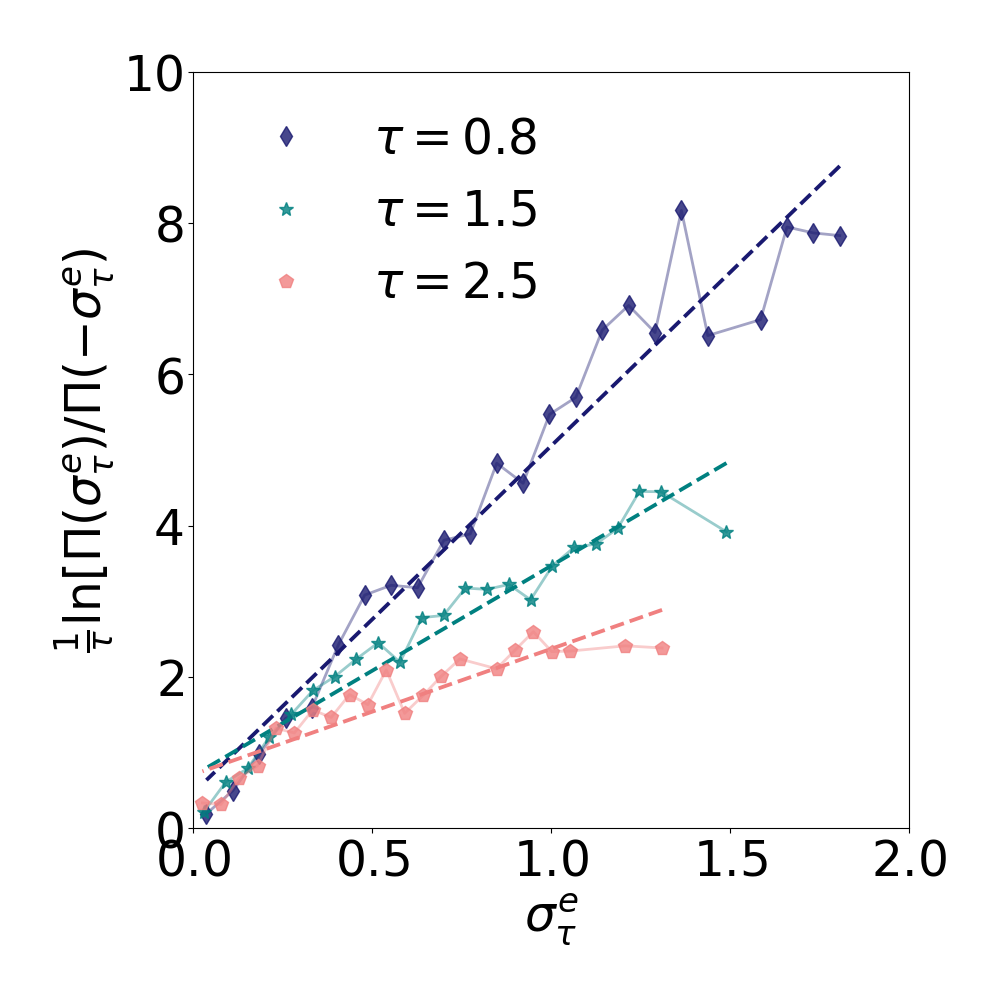}
		\put(-185,195){\bf{\large{(d)}}}
	\caption{Fluctuation relations for the BE. (a) PDFs of the energy injection rate, $\epsilon$, for the RB and BE in the quasi-equilibrium regime. (b) PDFs of viscosity $\nu_r$ for the RB in the quasi-equilibrium regime.  $s_{\epsilon}$ and $s_{\nu_r}$ are the standard deviations of $\epsilon$ and $\nu_r$, respectively. The black dashed-line represents the standard Gaussian PDF. (c) PDFs of $\sigma^{e}_{\tau}$, averaged over the interval $\tau$, for the BE (corresponds to $R=0.28$ for the RB).  (d) Plots to test the validity for FR in the BE for the energy injection rate, see Eq.~\eqref{eq:epr} and~\eqref{eq:GCFR}.}
	\label{fig:FR1}
\end{figure*}

In this section, the temporal fluctuations of certain global quantities of both the BE and RB are examined, and finally the fluctuation relation obeyed by them is introduced. For example, for the BE, $E$ fluctuates in time, whereas $\nu$ is constant by construction; but for the RB, it is $\nu_r$ which exhibits temporal fluctuations and $E$ is constant in time. For both the systems, the global quantities $\epsilon$ and $\Omega$ fluctuate in time; the latter remains positive definite. Interestingly, $\epsilon$ and $\nu_r$ can become negative at certain instants of time. Note that for the RB, $\nu_r = \epsilon/\Omega$, which suggests that the negative viscosity events are directly correlated with $\epsilon$. It must be emphasized that the negative viscosity values occur only in the quasi-equilibrium regime of the RB (see Ref.~\cite{Biferale_2018} for the 3D RNS). Moreover, in (a formally) time-reversible NSE, the fluctuating viscosity has been argued to be associated with the fluctuations of the entropy production rate~\cite{gallavotti1995dynamicalch,evans1993probability,aumaitre2001power}.

The GCFR in general deals with the phase space volume contraction rate $\Lambda(t)$, which in turn is related to the entropy production rate $\sigma^e$, for the time-symmetric RB system. Furthermore, the entropy production rate can be associated with the energy injection rate. The phase space contraction rate for the RB system is defined as
\begin{equation}
	\sigma^c = \Lambda(t)/\sqrt{\Var(\Lambda(t))},
\end{equation}
where $\Lambda$ is given by Eq.~(\ref{eq:pspRB}). Also, the first term on the right hand side of the Eq.~\eqref{eq:pspRB} is used to define a quantity analogous to the phase space contraction rate of the BE (see Eq.~\eqref{eq:pspBE}) as
\begin{equation}
	\Lambda^{\prime}(t) = \nu_r(t) \big[ \kmax(\kmax+1)(2\kmax+1)/3 \big].
\end{equation}
This is in turn used to define
\begin{equation}\label{eq:analogpsc}
	\sigma^{\prime \, c} = \Lambda^{\prime}(t)/\sqrt{\Var(\Lambda^{\prime}(t))}.
\end{equation}
Finally, the entropy production rate in terms of $\epsilon$ is given by
\begin{equation}\label{eq:epr}
	\sigma^e = \epsilon(t)/\sqrt{\Var(\epsilon(t))}.
\end{equation}

Observe that both $\epsilon/E_0$ and $\lambda(t)$ have the dimension of inverse time, but the above choices of $\sigma^e, (\sigma^{\prime})^c \text{ and }  \sigma^c$ renders them non-dimensional. A comparison of these three quantities in Fig.~\ref{fig:contraction} clearly shows a good agreement between them. Thus, this is indicative not only of an excellent agreement between the phase space contraction rate and the energy injection rate, but also that the first term in the definition of $\Lambda(t)$ is the dominant term in Eq.~\eqref{eq:pspRB} (for the RB). This further suggests an equivalence between the RB and BE.

Let $\Pi$ be the probability distribution of some time-fluctuating quantity $x$ averaged over a time interval $\tau$, i.e.,
\begin{equation}
	x_{\tau} = \frac{1}{\tau} \int^{\tau}_{0} x(t) \, \textrm{dt}.
\end{equation}
The FR for the quantity $x$ is given by
\begin{equation}\label{eq:GCFR}
	\lim_{\tau \to \infty} \frac{1}{\tau}\ln \frac{\Pi(x_{\tau}=p)}{\Pi(x_{\tau}=-p)}=\beta p.
\end{equation}
For a Gaussian distributed $x$, it can be shown that $\beta = 2 \langle x \rangle/\Var{ (x) }$ in the limit of a large time-interval $\tau$~\cite{Zamponi2004}. For $x=\sigma^c$, the above FR coincides with the GCFR.

The above GCFR is tested by computing the left hand side of the Eq.~\eqref{eq:GCFR}, with $x = \sigma^c$ for finite time intervals of increasing size and estimating the slope $\beta_{\tau}$. In the large time interval limit, $\tau \to \infty$, a convergence in the values of $\beta_{\tau}$, \textit{i.e.}, $\beta_{\tau} \to \beta$, would indicate that the GCFR holds. Note that while doing this, the decorrelation time of the observables involved must be kept in mind. Here, it is proportional to the time of appearance of the shock $t_{*}=1/u_0$, as the shock is the precursor to the thermalization. The value of shock time can be analytically derived for the unforced viscous BE and the inviscid BE (Appendix E of~\cite{das2023rb}). For small $R$, the decorrelation time is small; hence, the $\tau$ values considered here are much larger.

Figures~\ref{fig:FR} (a) and (b) show the probability distribution $\Pi$ and the L.H.S. of Eq.~\eqref{eq:GCFR} of the FR, respectively, for $\sigma^e_{\tau}$ for different values of $\tau$. Similarly, Fig.~\ref{fig:FR} (c)  and (d) show the $\Pi$ and L.H.S. of Eq.~\eqref{eq:GCFR}, respectively, for $\sigma^c_{\tau}$. The PDFs in Figures~\ref{fig:FR} (a) and (c) are approximately Gaussian, with positive mean, and become more peaked as the length of the time interval increases. Also, Fig.~\ref{fig:FR} (b) and (d) indicate that $(1/\tau)\ln[\Pi_{\tau}(p)/\Pi_{\tau}(-p)]$ is linear in $p$ for both the quantities. Moreover, it is clearly evident from Fig.~\ref{fig:betaconverge} that the slopes $\beta_{\tau}$ decrease as $\tau$ increases and tend to converge for large time intervals to the same limit. Observe that the convergence is much faster for the smaller values of $R$, as the decorrelation time is proportional to $t^* \propto R^{1/2}$ ($\sim 1/u_0$).

Figures~\ref{fig:FR1} (a) and (b) show the PDFs of fluctuations in the energy injection rate $\sigma^e$ and the reversible viscosity $\nu_r$, respectively, for different values of $R$. These PDFs are described very well by a Gaussian distribution. Note that Fig.~\ref{fig:FR1} (a) also shows that the PDFs of fluctuations in the energy injection rate for the BE and RB systems are in a perfect agreement with each other and collapse onto each other for different values of $R$ in the thermalized regime. Next the validity of the Cohen-Searles FRs for the BE system (corresponding to $R=0.28$ for the RB) is examined by using the quantity $\sigma^e$. Figure~\ref{fig:FR1} (c) shows the PDFs of $\sigma^e_{\tau}$ for different values of $\tau$. It is interesting to observe that a plot of $(1/\tau)\ln[\Pi_{\tau}(p)/\Pi_{\tau}(-p)]$ vs $p$ exhibits a linear regime, to a good approximation, thereby confirming the validity of the Cohen-Searles FR, see Fig.~\ref{fig:FR1} (d). Such a FR relation was examined using the 3D shell model of turbulence in~\cite{aumaitre2001power}.

The validation of the FR for the energy injection rate in case of Galerkin-truncated BE indicates that the thermalized states of the quasi-equilibrium regime follow time-reversible dynamics. Similarly, for the RB, the time-reversible dynamics is evident only in the quasi-equilibrium regime, as suggested by the FRs for both the energy injection rate and the phase-space contraction rate. The validation of the FR and the pairing symmetry in the Lyapunov exponents are linked, since the time-reversible nature of the dynamics depends on the properties of the attractor of the system. Also, it is interesting to see that both GCFR and the Cohen-Searles FRs hold true.

\section{Concluding Summary}

In this work, Galerkin-truncated BE and RB are revisited and the statistics of a quasi-equilibrium (thermalized) regime present in each is examined. The fluctuations of Fourier modes of the velocity are found to follow a Gaussian distribution. The distributions of other derived quantities considered are found using this result. The observed violations of the second-law, in this regime, are argued to be the result of these thermal noise fluctuations born out of the adverse truncation effects. The analytical expressions for the Jacobian matrices of both the BE and the RB are derived and the local Lyapunov spectra are computed for each in the hydrodynamic and quasi-equilibrium regimes. These results are found to be in an excellent agreement with those obtained from the DNSs.

A pairing symmetry is seen in the Lyapunov spectra in the quasi-equilibrium regime, which is absent in the hydrodynamic regime. The symmetry in the quasi-equilibrium regime suggests that the attractor of the dynamics has the same dimension as the entire phase space, which is not true in the hydrodynamic regime. In the shock dominated regime, sans truncation effects, the absence of pairing suggests that the dimension of the attractor $\mathcal{A}$ is less than that of the phase space. As the dynamics becomes confined to the attractor, the time-reversible symmetry does not remain viable, since the image of the attractor under the time-reversal symmetry, $\mathcal{IA}$ may not be on the attractor, \textit{i.e.}, $\mathcal{IA}\neq \mathcal{A}$~\cite{gallavotti2020nonequilibrium}. Therefore, even though the governing equations are time-reversible, the dynamics is effectively irreversible. Hence, the features associated with a reversible dynamics such as the events of violations of second-law become unlikely. Also, as a consequence the FRs become inadmissible. However, if the attractor spans the entire phase space, the likelihood of such events increases, as the reversible dynamics is restored. This is indeed what is observed.

The Lyapunov spectra can be used to define a `fluctuation dimension' of the attractor of the system that is equal to twice the number of non-negative Lyapunov exponents. Similarly, the Kaplan-Yorke dimension~\cite{evans2002fluctuation} is also used to study the fractal properties of the attractor. It is given by
\begin{equation}
D^{KY} = n^{KY} + \sum_{i} \frac{\lambda_i}{ | \lambda_{n^{KY}+1} |},
\end{equation}
where $\lambda$'s are the set of eigenvalues arranged in a descending fashion and $n^{KY}$ is the largest integer for which $ \sum_{i} \lambda_i > 0 $.
The pairing symmetry in the quasi-equilibrium regime suggests that $\mathcal{A}$ has the fluctuation dimension equal to that of the entire phase space. Also, it is found that $D^{KY} = 2\kmax+1$, in this regime, which is the same as the dimension of the phase space. The fluctuation dimension and the Kaplan-Yorke dimension both indicate that the attractor $\mathcal{A}$ spans the entire phase space in the quasi-equilibrium regime. In other regimes, both dimensions are fractions of the phase space dimension. All this appears due to pairing symmetry being restored. This is likely to happen since the truncated inviscid BE posses a Hamiltonian formulation~\cite{abramovmajda2003}.

The Jacobians are used to find expressions for phase space contraction rates of the BE and RB. It is found to be a constant term for the BE and a fluctuating quantity for the RB. The GCFR which deals phase space contraction rates is examined for the RB and it holds valid. Moreover, the energy injection rate is found to satisfy the Cohen-Searles FR in the quasi-equilibrium regime of both the RB and the BE. Therefore, the FRs, Lyapunov pairing and the negative events are all effects of thermalization taking root in the system.

\vspace{1cm}

\textbf{Acknowlegments} The authors acknowledge National Supercomputing Mission (NSM) for providing computing resources of ‘PARAM Shakti’ at IIT Kharagpur, which is implemented by C-DAC and supported by the Ministry of Electronics and Information Technology (MeitY) and Department of Science and Technology (DST), along with the NSM Grant DST/NSM/R\&D HPC Applications/2021/03.21.  VS would like to acknowledge support from the Institute Scheme for Innovative Research and Development (ISIRD), IIT Kharagpur, Grant No. IIT/SRIC/ISIRD/2021-2022/03.

\appendix

\section{Simulation details of the Galerkin-truncated BE and RB}
\label{app:dns}

The equations for the BE~(\ref{eqn: BE}) and RB~(\ref{eqn: RB}) are integrated using a pseudospectral method over a domain of length $2\pi$. The Fourier series, \[ u(x,t)=\sum_{k=-\infty}^{\infty} u(k,t)e^{ikx}\] is Galerkin truncated using the $2/3$-dealiasing rule. This sets the maximum number of Fourier modes in the simulation as $\kmax = N_c/3$, where $N_c$ is the number of collocation points. All the results depicted are simulated using $N_c=1024$ \textit{i.e.} $\kmax=341$. Time-stepping is done using the standard RK4 scheme using sufficiently small time steps to ensure good conservation of energy for the RB.

%\bibliographystyle{apsrev4-1}
%\bibliography{reference}

%merlin.mbs apsrev4-1.bst 2010-07-25 4.21a (PWD, AO, DPC) hacked
%Control: key (0)
%Control: author (72) initials jnrlst
%Control: editor formatted (1) identically to author
%Control: production of article title (-1) disabled
%Control: page (0) single
%Control: year (1) truncated
%Control: production of eprint (0) enabled
\begin{thebibliography}{39}%
\makeatletter
\providecommand \@ifxundefined [1]{%
 \@ifx{#1\undefined}
}%
\providecommand \@ifnum [1]{%
 \ifnum #1\expandafter \@firstoftwo
 \else \expandafter \@secondoftwo
 \fi
}%
\providecommand \@ifx [1]{%
 \ifx #1\expandafter \@firstoftwo
 \else \expandafter \@secondoftwo
 \fi
}%
\providecommand \natexlab [1]{#1}%
\providecommand \enquote  [1]{``#1''}%
\providecommand \bibnamefont  [1]{#1}%
\providecommand \bibfnamefont [1]{#1}%
\providecommand \citenamefont [1]{#1}%
\providecommand \href@noop [0]{\@secondoftwo}%
\providecommand \href [0]{\begingroup \@sanitize@url \@href}%
\providecommand \@href[1]{\@@startlink{#1}\@@href}%
\providecommand \@@href[1]{\endgroup#1\@@endlink}%
\providecommand \@sanitize@url [0]{\catcode `\\12\catcode `\$12\catcode
  `\&12\catcode `\#12\catcode `\^12\catcode `\_12\catcode `\%12\relax}%
\providecommand \@@startlink[1]{}%
\providecommand \@@endlink[0]{}%
\providecommand \url  [0]{\begingroup\@sanitize@url \@url }%
\providecommand \@url [1]{\endgroup\@href {#1}{\urlprefix }}%
\providecommand \urlprefix  [0]{URL }%
\providecommand \Eprint [0]{\href }%
\providecommand \doibase [0]{http://dx.doi.org/}%
\providecommand \selectlanguage [0]{\@gobble}%
\providecommand \bibinfo  [0]{\@secondoftwo}%
\providecommand \bibfield  [0]{\@secondoftwo}%
\providecommand \translation [1]{[#1]}%
\providecommand \BibitemOpen [0]{}%
\providecommand \bibitemStop [0]{}%
\providecommand \bibitemNoStop [0]{.\EOS\space}%
\providecommand \EOS [0]{\spacefactor3000\relax}%
\providecommand \BibitemShut  [1]{\csname bibitem#1\endcsname}%
\let\auto@bib@innerbib\@empty
%</preamble>
\bibitem [{\citenamefont {Lebowitz}(1993)}]{lebowitz1993boltzmann}%
  \BibitemOpen
  \bibfield  {author} {\bibinfo {author} {\bibfnamefont {J.~L.}\ \bibnamefont
  {Lebowitz}},\ }\href@noop {} {\bibfield  {journal} {\bibinfo  {journal}
  {Physics today}\ }\textbf {\bibinfo {volume} {46}},\ \bibinfo {pages} {32}
  (\bibinfo {year} {1993})}\BibitemShut {NoStop}%
\bibitem [{\citenamefont {Evans}\ \emph {et~al.}(1993)\citenamefont {Evans},
  \citenamefont {Cohen},\ and\ \citenamefont {Morriss}}]{evans1993probability}%
  \BibitemOpen
  \bibfield  {author} {\bibinfo {author} {\bibfnamefont {D.~J.}\ \bibnamefont
  {Evans}}, \bibinfo {author} {\bibfnamefont {E.~G.~D.}\ \bibnamefont {Cohen}},
  \ and\ \bibinfo {author} {\bibfnamefont {G.~P.}\ \bibnamefont {Morriss}},\
  }\href@noop {} {\bibfield  {journal} {\bibinfo  {journal} {Physical review
  letters}\ }\textbf {\bibinfo {volume} {71}},\ \bibinfo {pages} {2401}
  (\bibinfo {year} {1993})}\BibitemShut {NoStop}%
\bibitem [{\citenamefont {Gallavotti}\ and\ \citenamefont
  {Cohen}(1995{\natexlab{a}})}]{gallavotti1995dynamical}%
  \BibitemOpen
  \bibfield  {author} {\bibinfo {author} {\bibfnamefont {G.}~\bibnamefont
  {Gallavotti}}\ and\ \bibinfo {author} {\bibfnamefont {E.~G.~D.}\ \bibnamefont
  {Cohen}},\ }\href@noop {} {\bibfield  {journal} {\bibinfo  {journal}
  {Physical review letters}\ }\textbf {\bibinfo {volume} {74}},\ \bibinfo
  {pages} {2694} (\bibinfo {year} {1995}{\natexlab{a}})}\BibitemShut {NoStop}%
\bibitem [{\citenamefont {Jarzynski}(1997)}]{jarzynki1997freeenergy}%
  \BibitemOpen
  \bibfield  {author} {\bibinfo {author} {\bibfnamefont {C.}~\bibnamefont
  {Jarzynski}},\ }\href {\doibase 10.1103/PhysRevLett.78.2690} {\bibfield
  {journal} {\bibinfo  {journal} {Phys. Rev. Lett.}\ }\textbf {\bibinfo
  {volume} {78}},\ \bibinfo {pages} {2690} (\bibinfo {year}
  {1997})}\BibitemShut {NoStop}%
\bibitem [{\citenamefont {Kurchan}(1998)}]{kurchan1998fluctuation}%
  \BibitemOpen
  \bibfield  {author} {\bibinfo {author} {\bibfnamefont {J.}~\bibnamefont
  {Kurchan}},\ }\href@noop {} {\bibfield  {journal} {\bibinfo  {journal}
  {Journal of Physics A: Mathematical and General}\ }\textbf {\bibinfo {volume}
  {31}},\ \bibinfo {pages} {3719} (\bibinfo {year} {1998})}\BibitemShut
  {NoStop}%
\bibitem [{\citenamefont {Lebowitz}\ and\ \citenamefont
  {Spohn}(1999)}]{Lebowitz1999}%
  \BibitemOpen
  \bibfield  {author} {\bibinfo {author} {\bibfnamefont {J.~L.}\ \bibnamefont
  {Lebowitz}}\ and\ \bibinfo {author} {\bibfnamefont {H.}~\bibnamefont
  {Spohn}},\ }\href {\doibase 10.1023/A:1004589714161} {\bibfield  {journal}
  {\bibinfo  {journal} {Journal of Statistical Physics}\ }\textbf {\bibinfo
  {volume} {95}},\ \bibinfo {pages} {333} (\bibinfo {year} {1999})}\BibitemShut
  {NoStop}%
\bibitem [{\citenamefont {Chetrite}\ and\ \citenamefont
  {Gawedzki}(2008)}]{Chetrite2008}%
  \BibitemOpen
  \bibfield  {author} {\bibinfo {author} {\bibfnamefont {R.}~\bibnamefont
  {Chetrite}}\ and\ \bibinfo {author} {\bibfnamefont {K.}~\bibnamefont
  {Gawedzki}},\ }\href {\doibase 10.1007/s00220-008-0502-9} {\bibfield
  {journal} {\bibinfo  {journal} {Communications in Mathematical Physics}\
  }\textbf {\bibinfo {volume} {282}},\ \bibinfo {pages} {469} (\bibinfo {year}
  {2008})}\BibitemShut {NoStop}%
\bibitem [{\citenamefont {Hayashi}\ \emph {et~al.}(2010)\citenamefont
  {Hayashi}, \citenamefont {Ueno}, \citenamefont {Iino},\ and\ \citenamefont
  {Noji}}]{hayashi2010ft}%
  \BibitemOpen
  \bibfield  {author} {\bibinfo {author} {\bibfnamefont {K.}~\bibnamefont
  {Hayashi}}, \bibinfo {author} {\bibfnamefont {H.}~\bibnamefont {Ueno}},
  \bibinfo {author} {\bibfnamefont {R.}~\bibnamefont {Iino}}, \ and\ \bibinfo
  {author} {\bibfnamefont {H.}~\bibnamefont {Noji}},\ }\href {\doibase
  10.1103/PhysRevLett.104.218103} {\bibfield  {journal} {\bibinfo  {journal}
  {Phys. Rev. Lett.}\ }\textbf {\bibinfo {volume} {104}},\ \bibinfo {pages}
  {218103} (\bibinfo {year} {2010})}\BibitemShut {NoStop}%
\bibitem [{\citenamefont {Carberry}\ \emph {et~al.}(2004)\citenamefont
  {Carberry}, \citenamefont {Reid}, \citenamefont {Wang}, \citenamefont
  {Sevick}, \citenamefont {Searles},\ and\ \citenamefont
  {Evans}}]{carberry2004fluct}%
  \BibitemOpen
  \bibfield  {author} {\bibinfo {author} {\bibfnamefont {D.~M.}\ \bibnamefont
  {Carberry}}, \bibinfo {author} {\bibfnamefont {J.~C.}\ \bibnamefont {Reid}},
  \bibinfo {author} {\bibfnamefont {G.~M.}\ \bibnamefont {Wang}}, \bibinfo
  {author} {\bibfnamefont {E.~M.}\ \bibnamefont {Sevick}}, \bibinfo {author}
  {\bibfnamefont {D.~J.}\ \bibnamefont {Searles}}, \ and\ \bibinfo {author}
  {\bibfnamefont {D.~J.}\ \bibnamefont {Evans}},\ }\href {\doibase
  10.1103/PhysRevLett.92.140601} {\bibfield  {journal} {\bibinfo  {journal}
  {Phys. Rev. Lett.}\ }\textbf {\bibinfo {volume} {92}},\ \bibinfo {pages}
  {140601} (\bibinfo {year} {2004})}\BibitemShut {NoStop}%
\bibitem [{\citenamefont {Wang}\ \emph {et~al.}(2002)\citenamefont {Wang},
  \citenamefont {Sevick}, \citenamefont {Mittag}, \citenamefont {Searles},\
  and\ \citenamefont {Evans}}]{wang2002secondlawviol}%
  \BibitemOpen
  \bibfield  {author} {\bibinfo {author} {\bibfnamefont {G.~M.}\ \bibnamefont
  {Wang}}, \bibinfo {author} {\bibfnamefont {E.~M.}\ \bibnamefont {Sevick}},
  \bibinfo {author} {\bibfnamefont {E.}~\bibnamefont {Mittag}}, \bibinfo
  {author} {\bibfnamefont {D.~J.}\ \bibnamefont {Searles}}, \ and\ \bibinfo
  {author} {\bibfnamefont {D.~J.}\ \bibnamefont {Evans}},\ }\href {\doibase
  10.1103/PhysRevLett.89.050601} {\bibfield  {journal} {\bibinfo  {journal}
  {Phys. Rev. Lett.}\ }\textbf {\bibinfo {volume} {89}},\ \bibinfo {pages}
  {050601} (\bibinfo {year} {2002})}\BibitemShut {NoStop}%
\bibitem [{\citenamefont {Evans}\ and\ \citenamefont
  {Searles}(2002)}]{evans2002fluctuation}%
  \BibitemOpen
  \bibfield  {author} {\bibinfo {author} {\bibfnamefont {D.~J.}\ \bibnamefont
  {Evans}}\ and\ \bibinfo {author} {\bibfnamefont {D.~J.}\ \bibnamefont
  {Searles}},\ }\href@noop {} {\bibfield  {journal} {\bibinfo  {journal}
  {Advances in Physics}\ }\textbf {\bibinfo {volume} {51}},\ \bibinfo {pages}
  {1529} (\bibinfo {year} {2002})}\BibitemShut {NoStop}%
\bibitem [{\citenamefont {Sevick}\ \emph {et~al.}(2008)\citenamefont {Sevick},
  \citenamefont {Prabhakar}, \citenamefont {Williams},\ and\ \citenamefont
  {Searles}}]{annurev2008sevick}%
  \BibitemOpen
  \bibfield  {author} {\bibinfo {author} {\bibfnamefont {E.}~\bibnamefont
  {Sevick}}, \bibinfo {author} {\bibfnamefont {R.}~\bibnamefont {Prabhakar}},
  \bibinfo {author} {\bibfnamefont {S.~R.}\ \bibnamefont {Williams}}, \ and\
  \bibinfo {author} {\bibfnamefont {D.~J.}\ \bibnamefont {Searles}},\ }\href
  {\doibase https://doi.org/10.1146/annurev.physchem.58.032806.104555}
  {\bibfield  {journal} {\bibinfo  {journal} {Annual Review of Physical
  Chemistry}\ }\textbf {\bibinfo {volume} {59}},\ \bibinfo {pages} {603}
  (\bibinfo {year} {2008})}\BibitemShut {NoStop}%
\bibitem [{\citenamefont {Merhav}\ and\ \citenamefont
  {Kafri}(2010)}]{Merhav_2010}%
  \BibitemOpen
  \bibfield  {author} {\bibinfo {author} {\bibfnamefont {N.}~\bibnamefont
  {Merhav}}\ and\ \bibinfo {author} {\bibfnamefont {Y.}~\bibnamefont {Kafri}},\
  }\href {\doibase 10.1088/1742-5468/2010/12/P12022} {\bibfield  {journal}
  {\bibinfo  {journal} {Journal of Statistical Mechanics: Theory and
  Experiment}\ }\textbf {\bibinfo {volume} {2010}},\ \bibinfo {pages} {P12022}
  (\bibinfo {year} {2010})}\BibitemShut {NoStop}%
\bibitem [{\citenamefont {Evans}\ \emph {et~al.}(2005)\citenamefont {Evans},
  \citenamefont {Searles},\ and\ \citenamefont
  {Rondoni}}]{evans2005application}%
  \BibitemOpen
  \bibfield  {author} {\bibinfo {author} {\bibfnamefont {D.~J.}\ \bibnamefont
  {Evans}}, \bibinfo {author} {\bibfnamefont {D.~J.}\ \bibnamefont {Searles}},
  \ and\ \bibinfo {author} {\bibfnamefont {L.}~\bibnamefont {Rondoni}},\
  }\href@noop {} {\bibfield  {journal} {\bibinfo  {journal} {Physical Review
  E}\ }\textbf {\bibinfo {volume} {71}},\ \bibinfo {pages} {056120} (\bibinfo
  {year} {2005})}\BibitemShut {NoStop}%
\bibitem [{\citenamefont {Frisch}\ and\ \citenamefont
  {Bec}(2001)}]{frisch2001burgulence}%
  \BibitemOpen
  \bibfield  {author} {\bibinfo {author} {\bibfnamefont {U.}~\bibnamefont
  {Frisch}}\ and\ \bibinfo {author} {\bibfnamefont {J.}~\bibnamefont {Bec}},\
  }in\ \href@noop {} {\emph {\bibinfo {booktitle} {New trends in turbulence
  Turbulence: nouveaux aspects}}}\ (\bibinfo  {publisher} {Springer},\ \bibinfo
  {year} {2001})\ pp.\ \bibinfo {pages} {341--383}\BibitemShut {NoStop}%
\bibitem [{\citenamefont {Bec}\ and\ \citenamefont
  {Khanin}(2007)}]{bec2007burgers}%
  \BibitemOpen
  \bibfield  {author} {\bibinfo {author} {\bibfnamefont {J.}~\bibnamefont
  {Bec}}\ and\ \bibinfo {author} {\bibfnamefont {K.}~\bibnamefont {Khanin}},\
  }\href@noop {} {\bibfield  {journal} {\bibinfo  {journal} {Physics reports}\
  }\textbf {\bibinfo {volume} {447}},\ \bibinfo {pages} {1} (\bibinfo {year}
  {2007})}\BibitemShut {NoStop}%
\bibitem [{\citenamefont {Chowdhury}\ \emph {et~al.}(2000)\citenamefont
  {Chowdhury}, \citenamefont {Santen},\ and\ \citenamefont
  {Schadschneider}}]{chowdhury2000statistical}%
  \BibitemOpen
  \bibfield  {author} {\bibinfo {author} {\bibfnamefont {D.}~\bibnamefont
  {Chowdhury}}, \bibinfo {author} {\bibfnamefont {L.}~\bibnamefont {Santen}}, \
  and\ \bibinfo {author} {\bibfnamefont {A.}~\bibnamefont {Schadschneider}},\
  }\href@noop {} {\bibfield  {journal} {\bibinfo  {journal} {Physics Reports}\
  }\textbf {\bibinfo {volume} {329}},\ \bibinfo {pages} {199} (\bibinfo {year}
  {2000})}\BibitemShut {NoStop}%
\bibitem [{\citenamefont {Das}\ \emph {et~al.}(2024)\citenamefont {Das},
  \citenamefont {Dutta},\ and\ \citenamefont {Shukla}}]{das2023rb}%
  \BibitemOpen
  \bibfield  {author} {\bibinfo {author} {\bibfnamefont {A.}~\bibnamefont
  {Das}}, \bibinfo {author} {\bibfnamefont {P.}~\bibnamefont {Dutta}}, \ and\
  \bibinfo {author} {\bibfnamefont {V.}~\bibnamefont {Shukla}},\ }\href
  {\doibase 10.1103/PhysRevE.109.065108} {\bibfield  {journal} {\bibinfo
  {journal} {Phys. Rev. E}\ }\textbf {\bibinfo {volume} {109}},\ \bibinfo
  {pages} {065108} (\bibinfo {year} {2024})}\BibitemShut {NoStop}%
\bibitem [{\citenamefont {Gallavotti}(1996)}]{gallavotti1996equivalence}%
  \BibitemOpen
  \bibfield  {author} {\bibinfo {author} {\bibfnamefont {G.}~\bibnamefont
  {Gallavotti}},\ }\href@noop {} {\bibfield  {journal} {\bibinfo  {journal}
  {Physics Letters A}\ }\textbf {\bibinfo {volume} {223}},\ \bibinfo {pages}
  {91} (\bibinfo {year} {1996})}\BibitemShut {NoStop}%
\bibitem [{\citenamefont {Biferale}\ \emph {et~al.}(1998)\citenamefont
  {Biferale}, \citenamefont {Pierotti},\ and\ \citenamefont
  {Vulpiani}}]{biferale1998time}%
  \BibitemOpen
  \bibfield  {author} {\bibinfo {author} {\bibfnamefont {L.}~\bibnamefont
  {Biferale}}, \bibinfo {author} {\bibfnamefont {D.}~\bibnamefont {Pierotti}},
  \ and\ \bibinfo {author} {\bibfnamefont {A.}~\bibnamefont {Vulpiani}},\
  }\href@noop {} {\bibfield  {journal} {\bibinfo  {journal} {Journal of Physics
  A: Mathematical and General}\ }\textbf {\bibinfo {volume} {31}},\ \bibinfo
  {pages} {21} (\bibinfo {year} {1998})}\BibitemShut {NoStop}%
\bibitem [{\citenamefont {De~Pietro}\ \emph {et~al.}(2018)\citenamefont
  {De~Pietro}, \citenamefont {Biferale}, \citenamefont {Boffetta},\ and\
  \citenamefont {Cencini}}]{DePietro2018}%
  \BibitemOpen
  \bibfield  {author} {\bibinfo {author} {\bibfnamefont {M.}~\bibnamefont
  {De~Pietro}}, \bibinfo {author} {\bibfnamefont {L.}~\bibnamefont {Biferale}},
  \bibinfo {author} {\bibfnamefont {G.}~\bibnamefont {Boffetta}}, \ and\
  \bibinfo {author} {\bibfnamefont {M.}~\bibnamefont {Cencini}},\ }\href
  {\doibase 10.1140/epje/i2018-11655-2} {\bibfield  {journal} {\bibinfo
  {journal} {The European Physical Journal E}\ }\textbf {\bibinfo {volume}
  {41}},\ \bibinfo {pages} {48} (\bibinfo {year} {2018})}\BibitemShut {NoStop}%
\bibitem [{\citenamefont {Biferale}\ \emph {et~al.}(2018)\citenamefont
  {Biferale}, \citenamefont {Cencini}, \citenamefont {Pietro}, \citenamefont
  {Gallavotti},\ and\ \citenamefont {Lucarini}}]{Biferale_2018}%
  \BibitemOpen
  \bibfield  {author} {\bibinfo {author} {\bibfnamefont {L.}~\bibnamefont
  {Biferale}}, \bibinfo {author} {\bibfnamefont {M.}~\bibnamefont {Cencini}},
  \bibinfo {author} {\bibfnamefont {M.~D.}\ \bibnamefont {Pietro}}, \bibinfo
  {author} {\bibfnamefont {G.}~\bibnamefont {Gallavotti}}, \ and\ \bibinfo
  {author} {\bibfnamefont {V.}~\bibnamefont {Lucarini}},\ }\href {\doibase
  10.1103/physreve.98.012202} {\bibfield  {journal} {\bibinfo  {journal}
  {Physical Review E}\ }\textbf {\bibinfo {volume} {98}} (\bibinfo {year}
  {2018}),\ 10.1103/physreve.98.012202}\BibitemShut {NoStop}%
\bibitem [{\citenamefont {Shukla}\ \emph {et~al.}(2019)\citenamefont {Shukla},
  \citenamefont {Dubrulle}, \citenamefont {Nazarenko}, \citenamefont
  {Krstulovic},\ and\ \citenamefont {Thalabard}}]{2019phshuklaase}%
  \BibitemOpen
  \bibfield  {author} {\bibinfo {author} {\bibfnamefont {V.}~\bibnamefont
  {Shukla}}, \bibinfo {author} {\bibfnamefont {B.}~\bibnamefont {Dubrulle}},
  \bibinfo {author} {\bibfnamefont {S.}~\bibnamefont {Nazarenko}}, \bibinfo
  {author} {\bibfnamefont {G.}~\bibnamefont {Krstulovic}}, \ and\ \bibinfo
  {author} {\bibfnamefont {S.}~\bibnamefont {Thalabard}},\ }\href@noop {}
  {\bibfield  {journal} {\bibinfo  {journal} {Physical Review E}\ }\textbf
  {\bibinfo {volume} {100}},\ \bibinfo {pages} {043104} (\bibinfo {year}
  {2019})}\BibitemShut {NoStop}%
\bibitem [{\citenamefont {Jaccod}\ and\ \citenamefont
  {Chibbaro}(2021)}]{jaccod2021const}%
  \BibitemOpen
  \bibfield  {author} {\bibinfo {author} {\bibfnamefont {A.}~\bibnamefont
  {Jaccod}}\ and\ \bibinfo {author} {\bibfnamefont {S.}~\bibnamefont
  {Chibbaro}},\ }\href {\doibase 10.1103/PhysRevLett.127.194501} {\bibfield
  {journal} {\bibinfo  {journal} {Phys. Rev. Lett.}\ }\textbf {\bibinfo
  {volume} {127}},\ \bibinfo {pages} {194501} (\bibinfo {year}
  {2021})}\BibitemShut {NoStop}%
\bibitem [{\citenamefont {Margazoglou}\ \emph {et~al.}(2022)\citenamefont
  {Margazoglou}, \citenamefont {Biferale}, \citenamefont {Cencini},
  \citenamefont {Gallavotti},\ and\ \citenamefont
  {Lucarini}}]{margazoglou2022nonequilibrium}%
  \BibitemOpen
  \bibfield  {author} {\bibinfo {author} {\bibfnamefont {G.}~\bibnamefont
  {Margazoglou}}, \bibinfo {author} {\bibfnamefont {L.}~\bibnamefont
  {Biferale}}, \bibinfo {author} {\bibfnamefont {M.}~\bibnamefont {Cencini}},
  \bibinfo {author} {\bibfnamefont {G.}~\bibnamefont {Gallavotti}}, \ and\
  \bibinfo {author} {\bibfnamefont {V.}~\bibnamefont {Lucarini}},\ }\href@noop
  {} {\bibfield  {journal} {\bibinfo  {journal} {Physical Review E}\ }\textbf
  {\bibinfo {volume} {105}},\ \bibinfo {pages} {065110} (\bibinfo {year}
  {2022})}\BibitemShut {NoStop}%
\bibitem [{\citenamefont {Maji}\ \emph {et~al.}(2023)\citenamefont {Maji},
  \citenamefont {Eswaran}, \citenamefont {Ghosh}, \citenamefont
  {Seshasayanan},\ and\ \citenamefont {Shukla}}]{Shukla2dequivalence2023}%
  \BibitemOpen
  \bibfield  {author} {\bibinfo {author} {\bibfnamefont {M.}~\bibnamefont
  {Maji}}, \bibinfo {author} {\bibfnamefont {K.~S.}\ \bibnamefont {Eswaran}},
  \bibinfo {author} {\bibfnamefont {S.}~\bibnamefont {Ghosh}}, \bibinfo
  {author} {\bibfnamefont {K.}~\bibnamefont {Seshasayanan}}, \ and\ \bibinfo
  {author} {\bibfnamefont {V.}~\bibnamefont {Shukla}},\ }\href {\doibase
  10.1103/PhysRevE.108.015102} {\bibfield  {journal} {\bibinfo  {journal}
  {Phys. Rev. E}\ }\textbf {\bibinfo {volume} {108}},\ \bibinfo {pages}
  {015102} (\bibinfo {year} {2023})}\BibitemShut {NoStop}%
\bibitem [{\citenamefont {Cohen}\ and\ \citenamefont
  {Rondoni}(1998)}]{cohen1998note}%
  \BibitemOpen
  \bibfield  {author} {\bibinfo {author} {\bibfnamefont {E.}~\bibnamefont
  {Cohen}}\ and\ \bibinfo {author} {\bibfnamefont {L.}~\bibnamefont
  {Rondoni}},\ }\href@noop {} {\bibfield  {journal} {\bibinfo  {journal}
  {Chaos: An Interdisciplinary Journal of Nonlinear Science}\ }\textbf
  {\bibinfo {volume} {8}},\ \bibinfo {pages} {357} (\bibinfo {year}
  {1998})}\BibitemShut {NoStop}%
\bibitem [{\citenamefont
  {Gallavotti}(2020{\natexlab{a}})}]{gallavotti2020nonequilibrium}%
  \BibitemOpen
  \bibfield  {author} {\bibinfo {author} {\bibfnamefont {G.}~\bibnamefont
  {Gallavotti}},\ }\href@noop {} {\bibfield  {journal} {\bibinfo  {journal}
  {Journal of Statistical Physics}\ }\textbf {\bibinfo {volume} {180}},\
  \bibinfo {pages} {172} (\bibinfo {year} {2020}{\natexlab{a}})}\BibitemShut
  {NoStop}%
\bibitem [{\citenamefont {Gallavotti}\ \emph {et~al.}(2004)\citenamefont
  {Gallavotti}, \citenamefont {Rondoni},\ and\ \citenamefont
  {Segre}}]{GALLAVOTTI2004338}%
  \BibitemOpen
  \bibfield  {author} {\bibinfo {author} {\bibfnamefont {G.}~\bibnamefont
  {Gallavotti}}, \bibinfo {author} {\bibfnamefont {L.}~\bibnamefont {Rondoni}},
  \ and\ \bibinfo {author} {\bibfnamefont {E.}~\bibnamefont {Segre}},\ }\href
  {\doibase https://doi.org/10.1016/j.physd.2003.09.029} {\bibfield  {journal}
  {\bibinfo  {journal} {Physica D: Nonlinear Phenomena}\ }\textbf {\bibinfo
  {volume} {187}},\ \bibinfo {pages} {338} (\bibinfo {year} {2004})},\ \bibinfo
  {note} {microscopic Chaos and Transport in Many-Particle Systems}\BibitemShut
  {NoStop}%
\bibitem [{\citenamefont {Gallavotti}(2022)}]{gallavotti2022navier}%
  \BibitemOpen
  \bibfield  {author} {\bibinfo {author} {\bibfnamefont {G.}~\bibnamefont
  {Gallavotti}},\ }\href@noop {} {\bibfield  {journal} {\bibinfo  {journal}
  {arXiv preprint arXiv:2211.02961}\ } (\bibinfo {year} {2022})}\BibitemShut
  {NoStop}%
\bibitem [{\citenamefont {Gallavotti}(2021)}]{Gallavotti2021lyap}%
  \BibitemOpen
  \bibfield  {author} {\bibinfo {author} {\bibfnamefont {G.}~\bibnamefont
  {Gallavotti}},\ }\href {\doibase 10.1007/s10955-021-02830-1} {\bibfield
  {journal} {\bibinfo  {journal} {Journal of Statistical Physics}\ }\textbf
  {\bibinfo {volume} {185}},\ \bibinfo {pages} {21} (\bibinfo {year}
  {2021})}\BibitemShut {NoStop}%
\bibitem [{\citenamefont
  {Gallavotti}(2020{\natexlab{b}})}]{Gallavotti2020ensturbfr}%
  \BibitemOpen
  \bibfield  {author} {\bibinfo {author} {\bibfnamefont {G.}~\bibnamefont
  {Gallavotti}},\ }\href {\doibase 10.1140/epje/i2020-11961-0} {\bibfield
  {journal} {\bibinfo  {journal} {The European Physical Journal E}\ }\textbf
  {\bibinfo {volume} {43}},\ \bibinfo {pages} {37} (\bibinfo {year}
  {2020}{\natexlab{b}})}\BibitemShut {NoStop}%
\bibitem [{\citenamefont {Gallavotti}(2018)}]{Gallavotti2018quant}%
  \BibitemOpen
  \bibfield  {author} {\bibinfo {author} {\bibfnamefont {G.}~\bibnamefont
  {Gallavotti}},\ }\href {\doibase 10.1140/epjst/e2018-700096-x} {\bibfield
  {journal} {\bibinfo  {journal} {The European Physical Journal Special
  Topics}\ }\textbf {\bibinfo {volume} {227}},\ \bibinfo {pages} {217}
  (\bibinfo {year} {2018})}\BibitemShut {NoStop}%
\bibitem [{\citenamefont {Banerjee}(2019)}]{banerjee2019fractional}%
  \BibitemOpen
  \bibfield  {author} {\bibinfo {author} {\bibfnamefont {D.}~\bibnamefont
  {Banerjee}},\ }\href@noop {} {\bibfield  {journal} {\bibinfo  {journal} {The
  European Physical Journal B}\ }\textbf {\bibinfo {volume} {92}},\ \bibinfo
  {pages} {1} (\bibinfo {year} {2019})}\BibitemShut {NoStop}%
\bibitem [{\citenamefont {Dressler}(1988)}]{lyapairfrictionDressler1998}%
  \BibitemOpen
  \bibfield  {author} {\bibinfo {author} {\bibfnamefont {U.}~\bibnamefont
  {Dressler}},\ }\href {\doibase 10.1103/PhysRevA.38.2103} {\bibfield
  {journal} {\bibinfo  {journal} {Phys. Rev. A}\ }\textbf {\bibinfo {volume}
  {38}},\ \bibinfo {pages} {2103} (\bibinfo {year} {1988})}\BibitemShut
  {NoStop}%
\bibitem [{\citenamefont {Gallavotti}\ and\ \citenamefont
  {Cohen}(1995{\natexlab{b}})}]{gallavotti1995dynamicalch}%
  \BibitemOpen
  \bibfield  {author} {\bibinfo {author} {\bibfnamefont {G.}~\bibnamefont
  {Gallavotti}}\ and\ \bibinfo {author} {\bibfnamefont {E.~G.~D.}\ \bibnamefont
  {Cohen}},\ }\href@noop {} {\bibfield  {journal} {\bibinfo  {journal} {Journal
  of Statistical Physics}\ }\textbf {\bibinfo {volume} {80}},\ \bibinfo {pages}
  {931} (\bibinfo {year} {1995}{\natexlab{b}})}\BibitemShut {NoStop}%
\bibitem [{\citenamefont {Auma{\^\i}tre}\ \emph {et~al.}(2001)\citenamefont
  {Auma{\^\i}tre}, \citenamefont {Fauve}, \citenamefont {McNamara},\ and\
  \citenamefont {Poggi}}]{aumaitre2001power}%
  \BibitemOpen
  \bibfield  {author} {\bibinfo {author} {\bibfnamefont {S.}~\bibnamefont
  {Auma{\^\i}tre}}, \bibinfo {author} {\bibfnamefont {S.}~\bibnamefont
  {Fauve}}, \bibinfo {author} {\bibfnamefont {S.}~\bibnamefont {McNamara}}, \
  and\ \bibinfo {author} {\bibfnamefont {P.}~\bibnamefont {Poggi}},\
  }\href@noop {} {\bibfield  {journal} {\bibinfo  {journal} {The European
  Physical Journal B-Condensed Matter and Complex Systems}\ }\textbf {\bibinfo
  {volume} {19}},\ \bibinfo {pages} {449} (\bibinfo {year} {2001})}\BibitemShut
  {NoStop}%
\bibitem [{\citenamefont {Zamponi}\ \emph {et~al.}(2004)\citenamefont
  {Zamponi}, \citenamefont {Ruocco},\ and\ \citenamefont
  {Angelani}}]{Zamponi2004}%
  \BibitemOpen
  \bibfield  {author} {\bibinfo {author} {\bibfnamefont {F.}~\bibnamefont
  {Zamponi}}, \bibinfo {author} {\bibfnamefont {G.}~\bibnamefont {Ruocco}}, \
  and\ \bibinfo {author} {\bibfnamefont {L.}~\bibnamefont {Angelani}},\ }\href
  {\doibase 10.1023/B:JOSS.0000028072.34588.32} {\bibfield  {journal} {\bibinfo
   {journal} {Journal of Statistical Physics}\ }\textbf {\bibinfo {volume}
  {115}},\ \bibinfo {pages} {1655} (\bibinfo {year} {2004})}\BibitemShut
  {NoStop}%
\bibitem [{\citenamefont {Abramov}\ \emph {et~al.}(2003)\citenamefont
  {Abramov}, \citenamefont {Kovačič},\ and\ \citenamefont
  {Majda}}]{abramovmajda2003}%
  \BibitemOpen
  \bibfield  {author} {\bibinfo {author} {\bibfnamefont {R.~V.}\ \bibnamefont
  {Abramov}}, \bibinfo {author} {\bibfnamefont {G.}~\bibnamefont {Kovačič}},
  \ and\ \bibinfo {author} {\bibfnamefont {A.~J.}\ \bibnamefont {Majda}},\
  }\href {\doibase https://doi.org/10.1002/cpa.3032} {\bibfield  {journal}
  {\bibinfo  {journal} {Communications on Pure and Applied Mathematics}\
  }\textbf {\bibinfo {volume} {56}},\ \bibinfo {pages} {1} (\bibinfo {year}
  {2003})},\ \Eprint
  {http://arxiv.org/abs/https://onlinelibrary.wiley.com/doi/pdf/10.1002/cpa.3032}
  {https://onlinelibrary.wiley.com/doi/pdf/10.1002/cpa.3032} \BibitemShut
  {NoStop}%
\end{thebibliography}%

%merlin.mbs apsrev4-1.bst 2010-07-25 4.21a (PWD, AO, DPC) hacked
%Control: key (0)
%Control: author (72) initials jnrlst
%Control: editor formatted (1) identically to author
%Control: production of article title (-1) disabled
%Control: page (0) single
%Control: year (1) truncated
%Control: production of eprint (0) enabled
%

\end{document}